\documentclass[aps,prb,twocolumn,superscriptaddress,amsmath,amssymb,eqsecnum]{revtex4-2}

\usepackage[colorlinks=true,linkcolor=blue,urlcolor=black,hyperfootnotes=true,citecolor=blue]{hyperref}
\usepackage{epsfig}
\usepackage{amsfonts}
\usepackage{braket}
\usepackage{epstopdf}
\usepackage{hyperref}
\usepackage{float}
\usepackage{bibentry}
\usepackage[caption=false]{subfig}
\usepackage{mathrsfs}
\usepackage{graphicx}
\usepackage{dcolumn}
\usepackage{bm}
\usepackage{color}

\begin{document}

\title{Field-resilient superconductivity in atomic-layer crystalline materials}

\author{Yoichi Higashi}
\email{y.higashi@aist.go.jp}
\affiliation{National Institute of Advanced Industrial Science and Technology (AIST), Tsukuba, Ibaraki 305-8568, Japan}

\author{Shunsuke Yoshizawa}
\affiliation{Center for Basic Research on Materials, National Institute for Materials Science (NIMS), 1-2-1 Sengen, Tsukuba, Ibaraki 305-0047, Japan.}

\author{Takashi Yanagisawa}
\affiliation{National Institute of Advanced Industrial Science and Technology (AIST), Tsukuba, Ibaraki 305-8568, Japan}

\author{Izumi Hase}
\affiliation{National Institute of Advanced Industrial Science and Technology (AIST), Tsukuba, Ibaraki 305-8568, Japan}

\author{Yasunori Mawatari}
\affiliation{National Institute of Advanced Industrial Science and Technology (AIST), Tsukuba, Ibaraki 305-8568, Japan}

\author{Takashi Uchihashi}
\affiliation{International Center for Materials Nanoarchitectonics (MANA), National Institute for Materials Science (NIMS), 1-1 Namiki, Tsukuba, Ibaraki 305-0044, Japan}
\date{\today}
\begin{abstract}
A recent study [S. Yoshizawa {\it et al}., Nature Communications {\bf 12}, 1462 (2021)] 
reported the occurrence of field-resilient superconductivity, that is, enhancement of the in-plane critical magnetic field $H^{||}_{\rm c2}$ beyond the paramagnetic limiting field,
in atomic-layer crystalline ($\sqrt{7}\times\sqrt{3}$)-In on a Si(111) substrate.
The present article elucidates the origin of the observed field-resilient noncentrosymmetric superconductivity in this highly crystalline two-dimensional material.
We develop the quasiclassical theory of superconductivity by incorporating the Fermi surface anisotropy
together with an anisotropic spin splitting and texture specific to atomic-layer crystalline systems. 
%
In Si(111)-($\sqrt{7}\times\sqrt{3}$)-In, a typical material with a large antisymmetric spin-orbit coupling (ASOC),
we show an example where the combination of the ASOC and disorder effect suppresses the paramagnetic depairing and can lead to an enhancement of $H^{||}_{\rm c2}$ compared to an isotropic system
only when a magnetic field is applied in a particular direction due to an anisotropic spin texture.
We also study the parity-mixing effect to demonstrate that the enhancement of $H^{||}_{\rm c2}$ is limited in the moderately clean regime
because of the fragile $s$+$p$-wave pairing
against nonmagnetic scattering in the case of the dominant odd-parity component of a pair wavefunction. 
Furthermore, from analysis of the transition line,
we identify the field-resilience factor taking account of the scattering and suppression of paramagnetic effects
and discuss the origin of the field-resilient superconductivity.
Through fitting of the $H^{||}_{\rm c2}$ data,
the normal-state electron scattering is discussed
with a prime focus on the role of atomic steps on a Si(111) surface. 

\end{abstract}

\maketitle
\section{Introduction}
Highly crystalline atomic-layer materials are currently attracting significant research interest as a new phase of matter associated with two-dimensional (2D) systems \cite{saito2016,uchihashi2016,gruznev2017review}.
The past decade has witnessed rapid progress in microfabrication technologies and atomic-layer materials research,
as well as the development and integration of measurement techniques applicable at ultralow temperatures and ultrahigh vacuums.
These advances have integrated superconductivity (SC) research and surface science,
enabling the exploration of 2D SC by measuring the superconducting properties of highly crystalline atomic-layer materials.


Previous studies into 2D SC have been extensively conducted using ultrathin amorphous or highly disordered metal films \cite{strongin1970,jaeger1986,haviland1989,sekihara2013,sekihara2015}.
In contrast to these systems, highly crystalline metal or alloy atomic-layer systems are fabricated on the reconstructions of semiconductor surfaces.
The single-atomic-layer SC of Pb and In epitaxially grown on Si(111) surfaces has been observed by scanning tunneling spectroscopy \cite{zhang2010}.
Furthermore, a robust supercurrent was observed on a macroscopic scale by electron transport measurements on a reconstruction of the Si(111) surface
with adsorbed In atoms [Si(111)--($\sqrt{7}\times\sqrt{3}$)-In] \cite{uchihashi2011}.
Subsequently, the SC of the single-atomic-layer alloy Si(111)--($\sqrt{3}\times\sqrt{3})$-Tl,Pb was confirmed by electron transport measurements \cite{matetskiy2015}.
Recently, the diamagnetic response of SC for Si(111)--($\sqrt{7}\times\sqrt{3}$)-In was reported as well \cite{wu2019}.
Herein, the $\sqrt{7}\times\sqrt{3}$ or $\sqrt{3}\times\sqrt{3}$ indicates the enlarged ratio of the surface superstructure unit cell to the bulk silicon crystal one.

Because the spatial inversion symmetry is intrinsically broken on a semiconductor substrate surface,
heavy-element atomic layers on top of substrates accommodate spin-split energy bands owing to spin--orbit coupling (SOC) \cite{petersen2000,friedel1964,meservey1976,uchihashi2021},
which is referred to as antisymmetric SOC (ASOC) in the context of noncentrosymmetric SC \cite{frigeri2004njp}.
Hence, the Fermi surface (FS) is allowed to split into two by lifting the spin degeneracy,
and parity mixing of the pair wavefunction must occur, although its realization depends on the material parameters \cite{anderson1984,gorkov2001,frigeri2004}.
Indeed, spin-split metallic energy bands in the normal state have been observed by angle-resolved photoelectron spectroscopy (ARPES)
for Si(111)--($\sqrt{3}\times\sqrt{3})$-Tl,Pb near the Fermi energy $\varepsilon_{\rm F}$,
with maximum energy splitting widths of 250 and 140 meV ($\sim 10^3|\varDelta|$) for the $\varSigma_1$ and $\varSigma_2$ bands, respectively \cite{gruznev2014,matetskiy2015}, where $|\varDelta|$ denotes the superconducting gap. 
Regarding Si(111)--($\sqrt{7}\times\sqrt{3}$)-In,
while the FS was observed by ARPES, spin-split energy bands were not because of the limited momentum resolution \cite{rotenberg2003}.
Recently, spin polarization on the butterfly-shaped FS was confirmed by spin-resolved ARPES,
and the spin texture on the FS suggests a non-ideal Rashba ASOC associated with the lower point group symmetry $C_{\rm 1h}$ at the surface \cite{kobayashi2020}.
The maximum observed energy splitting at $\varepsilon_{\rm F}$ is 87 meV ($\sim 10^2|\varDelta|$), which is consistent with the results of density functional theory (DFT) calculations \cite{yoshizawa2021,kobayashi2020}.

Meanwhile, the magnetic properties of atomic-layer SC have also been revealed.
Suppression of paramagnetic depairing due to the Zeeman-type ASOC in Ising superconductors such as MoS$_2$ leads to significant enhancement of the in-plane critical field at zero temperature $H^{||}_{\rm c2}(T\approx 0)$, giving a maximum value in excess of 50 T \cite{saito2016natphys}.
For isotropic Rashba ASOC \cite{rashba1960,bychkov1984}, $H^{||}_{\rm c2}$ is limited to $\sqrt{2}H^{\rm P}$ \cite{bulaevskii1973}, where $H^{\rm P}$ is the conventional Pauli limiting field. 
However, experimental results suggest $H^{||}_{\rm c2}$ values well above $H^{\rm P}$ for Si(111)--$(\sqrt{7}\times\sqrt{3})$-In \cite{yoshizawa2021}.

Investigation of the influence of Pauli paramagnetism on the magnetic properties was pioneered by Maki,
who scrutinized the role of spin--orbit scattering \cite{maki1966}.  
Several theoretical studies have elucidated the $T$--$H$ phase diagram for isotropic Rashba SC in the strong SOC regime to demonstrate the $H^{||}_{\rm c2}$ enhancement
upon increasing the density of Born scatterers \cite{dimitrova2003,dimitrova2007,samokhin2008}
or introducing helical and stripe modulations \cite{agterberg2007,yanase2008,zwicknagl2017}.
For an arbitrary SOC strength, the enhancement of $H^{||}_{\rm c2}$ under a helical modulation is shown in the dirty limit \cite{houzet2015}.

While most previous theories have been applied to isotropic systems \cite{bychkov1984,gorkov2001,barzykin2002,dimitrova2003,frigeri2004,kaur2005,frigeri2006,hayashi2006,dimitrova2007,agterberg2007,samokhin2008,houzet2015,higashi2016,zwicknagl2017},
there are limited examples of their application to anisotropic systems such as bulk noncentrosymmetric crystals \cite{yanase2007,goryo2012,youn2012}, oxide-heterostructure interfaces \cite{michaeli2012,nakamura2013}, and highly crystalline atomic layer materials \cite{saito2016natphys,nakamura2017}.
Highly crystalline atomic-layer materials possess anisotropic FSs that have ever been clearly observed with a spin texture structure.
Because highly disordered alloys and amorphous thin films lack an anisotropic FS,
this is an important feature for understanding the paramagnetic properties of atomic-layer SC.
Furthermore, spin texture is not incorporated into the equation to determine $H^{||}_{\rm c2}$ \cite{maki1966,klemm1975}.
Therefore, the conventional isotropic description is unsatisfactory for explaining the $H^{||}_{\rm c2}$ enhancement observed in highly crystalline atomic-layer superconductors.

In this study, by incorporating the anisotropic FS with spin texture obtained by DFT calculations and the parity-mixing effect, we develop the quasiclassical theory of SC by extending it to highly crystalline 2D superconductors.
We apply the developed theory to the SC in Si(111)--$(\sqrt{7}\times\sqrt{3})$-In to demonstrate the enhancement of $H^{||}_{\rm c2}(T\approx 0)$ to above $\sqrt{2}H^{\rm P}$, that is, the occurrence of magnetic-field resilience.
The enhancement of $H^{||}_{\rm c2}(T\approx 0)$ compared to an isotropic system results from the combination of the ASOC and disorder effect.
It is not always present, but depends on the in-plane field direction due to an anisotropic spin texture.
We also compare the numerical results with the available experimental data for $H^{||}_{\rm c2}(T\approx T_{\rm c})$ in Si(111)--$(\sqrt{7}\times\sqrt{3})$-In for vicinal substrates  
and discuss the normal-state electron scattering off atomic steps.

The remainder of this paper is organized as follows.
In Sec. \ref{sec:II}, we present the anisotropic FSs and spin texture on the butterfly-shaped FSs obtained by DFT calculations for Si(111)--$(\sqrt{7}\times\sqrt{3})$-In.
Sec. \ref{sec:III} is devoted to the self-consistent equations based on the quasiclassical theory in the strong SOC regime, which is applicable to highly crystalline 2D atomic-layer materials.
In Sec. \ref{sec:IV}, we examine a parallel critical field and discuss the origin of the field resilience in Si(111)--$(\sqrt{7}\times\sqrt{3})$-In,
and we also compare the numerical results for $H^{||}_{\rm c2}$ with the experimental data to estimate the normal-state electron scattering rate.
In Sec. \ref{sec:V}, the possible orbital effect and normal-state electron scattering are discussed with a focus on the role of atomic steps. 
Finally, a brief summary is provided in Sec. \ref{sec:V}.
\section{Anisotropic spin splitting and spin texture}
\label{sec:II}
\begin{figure}[tb]
  \begin{center}
    \begin{tabular}{p{85mm}p{85mm}}
      \resizebox{85mm}{!}{\includegraphics{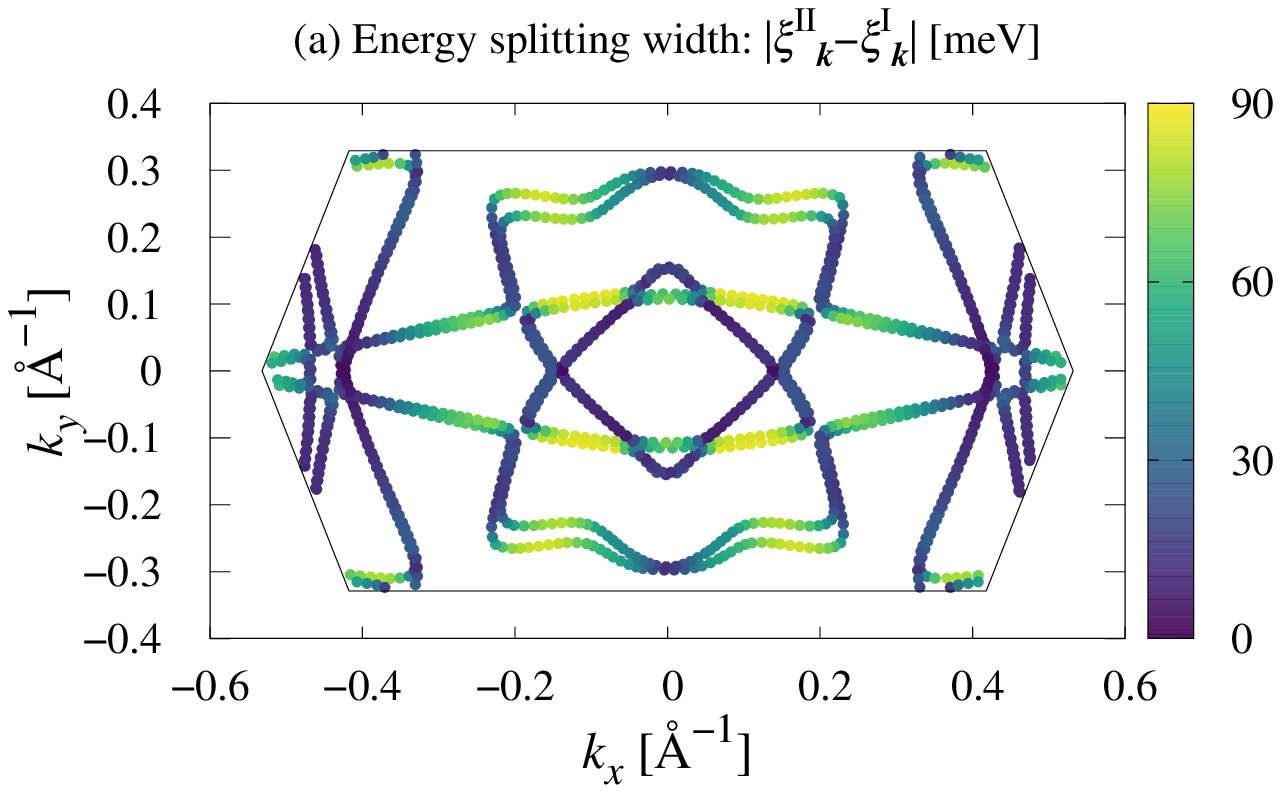}}\\
      \resizebox{85mm}{!}{\includegraphics{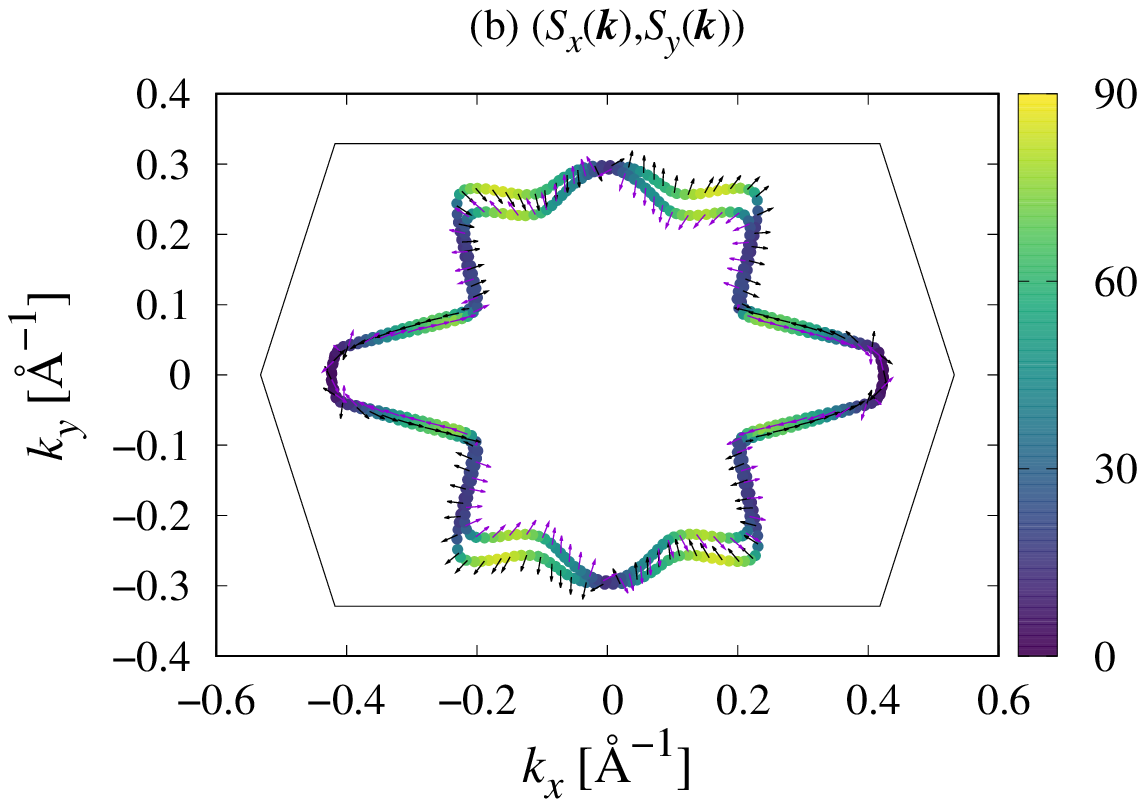}}
    \end{tabular}
\caption{
\label{fig1}
(Color online) (a) Anisotropic Fermi surface of Si(111)--($\sqrt{7}\times\sqrt{3})$-In, where the color scale indicates the energy splitting width  $|\xi^{\rm II}_{\bm{k}}-\xi^{\rm I}_{\bm{k}}|$.
(b) Spin texture $(S_x(\bm{k}),S_y(\bm{k}))$ on the FSs with DOSs an order of magnitude larger than those for the other FSs.
The solid line indicates the first Brillouin zone.
}
  \end{center}
\end{figure}
\begin{center}
\begin{table}[t]
\caption{Densities of states (DOSs) at the Fermi level.}
\begin{tabular}{ccc}
\hline
\hline
Band index  & DOS $[{\rm \AA}^{-2}{\rm eV}^{-1}]$ & Proportion [\%]\\
\hline
396 & 5.46 $\times 10^{-3}$ &5.92\\
397 & 5.35 $\times 10^{-3}$ &5.80\\
398 & 9.45 $\times 10^{-3}$ &10.3\\
399 & 7.56 $\times 10^{-3}$ &8.20\\
400 & 2.33 $\times 10^{-2}$ &25.2\\
401 & 2.07 $\times 10^{-2}$ &22.4\\
402 & 6.92 $\times 10^{-3}$ &7.51\\
403 & 6.14 $\times 10^{-3}$ &6.66\\
404 & 3.61 $\times 10^{-3}$ &3.91\\
405 & 3.72 $\times 10^{-3}$ &4.04\\
\hline
\hline
\end{tabular}
\label{table1}
\end{table}
\end{center}

Figure~\ref{fig1} depicts the energy splitting width and spin texture on the FS within the Brillouin zone for Si(111)--$(\sqrt{7}\times\sqrt{3})$-In, obtained by DFT calculations with SOC.
These calculations were performed with the Quantum ESPRESSO suite of codes \cite{giannozzi2009}.
We employed the augmented plane wave method, and used the local density approximation for the exchange correlation.
The crystal structure was modeled by a repeated slab, and the geometry optimization was performed without including the SOC.
The resulting atomic coordinates show good agreement with diffraction measurements \cite{shirasawa2019}.
Further details on the computational conditions can be found in Refs.~\cite{yoshizawa2021,shirasawa2019}.

The spin-split bands due to the SOC are denoted $\xi^{\rm I,II}_{\bm k}$, as described in more detail in Sec.~\ref{sec:III}.
As shown in Table \ref{table1},
the densities of states (DOSs) at $\varepsilon_{\rm F}$ for bands 400 and 401 are an order of magnitude larger than those for the other FSs.
The proportions of the DOSs at $\varepsilon_{\rm F}$ to the total DOSs are 25.2\% and 22.4\% for bands 400 and 401, respectively.
Thus, we focus on a single pair of the split FSs originating from bands 400 and 401 as shown in Fig.~\ref{fig1}(b).
This simplification is reasonable because the selected FSs dominate the superconducting properties,
while the other FSs have an order of magnitude fewer states per unit energy contributing to the SC.
The arrows on the FSs in Fig.~\ref{fig1}(b) indicate the in-plane spin components $(S_x({\bm k}),S_y({\bm k}))$. 

Spin-split FSs in Rashba systems shift in an in-plane field,
resulting in phase modulation of the pair wavefunction in space in the presence of a DOS difference between the split FSs ({\it i.e.}, helical phase).
The modulation wavenumber is evaluated via $q=2\delta q_0$ with $q_0=m^\ast \mu_{\rm B} H/\hbar v_{\rm F}$ \cite{kaur2005} and $\delta=(N_{\rm I}-N_{\rm II})/2N_0\approx 0.0589$ the DOS weighting factor for the FSs in Fig.~\ref{fig1}(b).
$N_{\rm I,II}$ are the DOSs at $\varepsilon_{\rm F}$ for the spin-split bands and $N_0=(N_{\rm I}+N_{\rm II})/2$.
Correspondingly, the wavelength at 1 T is estimated as $\lambda={2\pi}/{q}\sim 867~{\rm \mu m}$ by adopting $m^\ast=1.14 m_{\rm e}$ and $\varepsilon_{\rm F}=6.60$ eV in the free electron model,
which gives rise to the material parameters $k_{\rm F}=1.41$ {\rm \AA}$^{-1}$ and $v_{\rm F}=1.43\times 10^6$ m/s obtained by DFT calculations for Si(111)--$(\sqrt{7}\times\sqrt{3})$-In.
The estimated $\lambda$ is several hundreds of micrometers at several tesla, which is comparable to the sample size.
Therefore, we neglect the helical modulation and focus instead on a uniform state in the long-wavelength limit to evaluate $H^{||}_{\rm c2}$.
This uniform state may survive against scatterers such as atomic defects in ($\sqrt{7}\times\sqrt{3}$)-In or the atomic steps inherent to a Si(111) substrate surface.

\section{Self-consistent equations in the strong spin-orbit coupling limit}
\label{sec:III}
We start with the normal-state Hamiltonian for 2D Rashba systems exposed to an in-plane field $\bm{H}$, 
\begin{equation}
\hat{\mathcal{H}}_{\rm n}=\xi_{\bm{k}}\hat{\sigma}_0+\left[ \bm{\lambda}_{\bm{k}}+\mu_{\rm B}\bm{H}\right]\cdot\hat{\bm{\sigma}},
\label{normal-hamiltonian}
\end{equation}
where $\xi_{\bm{k}}\equiv \varepsilon_{\bm{k}}-\mu$ is the electron band energy measured from the chemical potential $\mu$,
$\bm{\lambda}_{\bm{k}}$ is the vector characterizing the ASOC in energy units,
$\mu_{\rm B}$ is the Bohr magneton, and $\hat{\bm{\sigma}}=(\hat{\sigma}_x,\hat{\sigma}_y,\hat{\sigma}_z)^\top$ is the vector of the Pauli spin matrices.
The spin quantization axis is parallel to $\bm{H}$.
The vector potential is disregarded because of the quenched orbital motion of electrons in the present configuration.
Throughout the paper, $\hat{\cdot}$ denotes the $2\times2$ matrix in spin space.
From now on, we set $\hbar=k_{\rm B}=1$.

By transforming Eq.~(\ref{normal-hamiltonian}) into the Hamiltonian in the band basis where $\hat{\mathcal{H}}_{\rm n}$ in the absence of a magnetic field is diagonal,
we obtain the eigenenergy for each band split owing to the ASOC:
\begin{equation}
E^{\rm I,II}_{\bm{k}}\approx \xi^{\rm I,II}_{\bm{k}}\pm \mu_{\rm B} \bar{\bm{\lambda}}_{\bm{k}}\cdot\bm{H},
\label{excitation-ene-normal}
\end{equation}
where $\xi^{\rm I,II}_{\bm{k}}\equiv \xi_{\bm{k}}\pm |\bm{\lambda}_{\bm{k}}|$ and $\bar{\bm{\lambda}}_{\bm{k}}\equiv\bm{\lambda}_{\bm{k}}/|\bm{\lambda}_{\bm{k}}|$.
If we neglect the off-diagonal components describing the interband scattering induced by the in-plane field,
the model reduces to the effective two-band model.
By diagonalizing Eq.~(\ref{normal-hamiltonian}),
we also obtain the excitation energy [Eq.~(\ref{excitation-ene-normal})] up to the first order of $|\mu_{\rm B}\bm{H}|/|\bm{\lambda}_{\bm{k}}|$.
Herein, we assume $|\mu_{\rm B}\bm{H}|/|\bm{\lambda}_{\bm{k}}| \ll 1$, which is met for the condition of atomic-layer superconductors with a sufficiently large ASOC, $|\bm{\lambda}_{\bm{k}}|\gg T_{\rm c}$.
The condition $|\mu_{\rm B}\bm{H}|/|\bm{\lambda}_{\bm{k}}| \ll 1$ justifies the incorporation of the Zeeman field into the quasiclassical theory as a perturbation \cite{klein2000,ichioka2007,higashi2014,dan2015,higashi2016}.

In the effective two-band model,
we phenomenologically view the vector characterizing the ASOC to be defined for each band; $\bm{\lambda}_{\bm{k}}\rightarrow \bm{\lambda}^l_{\bm{k}}$ with $l=$ I or II denoting the band index.
The ASOC is characterized through $\bm{\lambda}^l_{\bm{k}}=\sqrt{ \langle |\bm{\lambda}^l_{\bm{k}}|^2\rangle_{\bm{k}} }\bm{g}^l_{\bm{k}}$
by the antisymmetric orbital vector
\begin{equation}
\bm{g}^l_{\bm{k}} = \cfrac{|\Delta\xi^l_{\bm{k}}|}{\sqrt{\left\langle |\Delta\xi^l_{\bm{k}}|^2\right\rangle_0}}\cfrac{\bm{S}^l_{\bm{k}}}{|\bm{S}^l_{\bm{k}}|},
\label{orbital-vector}
\end{equation}
which is set to be normalized as $\left\langle (\bm{g}^l_{{\bm{k}}})^2\right\rangle_0=1$ in accordance with an isotropic Rashba system.
In Eq.~(\ref{orbital-vector}), $|\Delta\xi^l_{\bm{k}}|/\sqrt{\left\langle |\Delta\xi^l_{\bm{k}}|^2\right\rangle_0}$ represents the anisotropy of the spin-split energy relative to
the typical energy scale of the ASOC.
Here, $\left\langle \cdots\right\rangle_0$ denotes an average over the FS in the absence of the ASOC and $\bm{H}$:
$\langle \cdots \rangle_0 \equiv \int{\rm d}S^{l}_{\rm F0}({\bm{k}})|\bm{v}^l_{\rm F0}({\bm{k}})|^{-1} \cdots /\int{\rm d}S^l_{\rm F0}({\bm{k}})|\bm{v}^l_{\rm F0}({\bm{k}})|^{-1}$
with ${\rm d}S^l_{\rm F0}({\bm{k}})$ and $\bm{v}^l_{\rm F0}({\bm{k}})$ the infinitesimal line element of the 2D FS and the Fermi velocity in the absence of the ASOC and $\bm{H}$, respectively. 
$|\Delta\xi^l_{\bm{k}}|\equiv |\xi^{\rm I}_{\bm{k}}-\xi^{\rm II}_{\bm{k}}|$ is the energy splitting width.
The maximum spin-split energy is $|\Delta \xi^{400}|\approx 84~{\rm meV}$ and $|\Delta \xi^{401}|\approx 80~{\rm meV}$
for the bands 400 and 401, respectively.
By adopting experimental value of the transition temperature $T^{\rm vicinal}_{\rm c}(0)\approx3.05~{\rm K}$ at zero field
for ($\sqrt{7}\times\sqrt{3}$)-In on a vicinal Si(111) surface,
the maximum value of ASOC is estimated as $|\Delta \xi^{400}|/T^{\rm vicinal}_{\rm c}(0)\approx 320$ and $|\Delta \xi^{401}|/T^{\rm vicinal}_{\rm c}(0)\approx 302$ for the bands 400 and 401, respectively.
The momentum-dependent spin polarization vector is obtained as in Fig.~\ref{fig1}(b) through $\bm{S}^l_{\bm{k}}=(1/2)\langle \Psi^l_{\bm{k}} |\hat{\bm{\sigma}} | \Psi^l_{\bm{k}}\rangle$ from the DFT calculations,
where $|\Psi^l_{\bm{k}}\rangle$ is the eigenstate of Eq.~(\ref{normal-hamiltonian}) at $\bm{H}=\bm{0}$.

The parity-mixed superconducting order parameters are determined via the self-consistent equations (\ref{gap-eq1_isotropic}) and (\ref{gap-eq2_isotropic}),
which are suitable for equivalent FSs ({\it i.e.}, infinitesimally split FSs) in the weak ASOC limit as proposed in Ref. \cite{hayashi2006}.
However, in the strong ASOC limit, the significantly split FSs are no longer equivalent.
Consequently, the average on the FS should be taken for each FS.
Thus, Eqs. (\ref{gap-eq1_isotropic}) and (\ref{gap-eq2_isotropic}) are recast as
\begin{align}
\psi_{\rm s}&=
\frac{\pi T}{2}\sum_{n=-[n_{\rm c}(T)]-1}^{[n_{\rm c}(T)]}
\big[
\lambda_{\rm s}\bigl\{ (1+\delta)\langle f_{\rm I}\rangle_{\rm I}+(1-\delta)\langle f_{\rm II}\rangle_{\rm II} \bigr\} \nonumber \\
&+\lambda_{\rm m}\bigl\{ (1+\delta)\langle |\bm{g}^{\rm I}_{\tilde{\bm{k}}}|f_{\rm I} \rangle_{\rm I} -(1-\delta)\langle |\bm{g}^{\rm II}_{\tilde{\bm{k}}}|f_{\rm II} \rangle_{\rm II}\bigr\}
 \big],
 \label{gap-eq1_insi}
 \\
d_{\rm t}&=
 \frac{\pi T}{2}\sum_{n=-[n_{\rm c}(T)]-1}^{[n_{\rm c}(T)]}
\big[
\lambda_{\rm m}\bigl\{ (1+\delta)\langle f_{\rm I}\rangle_{\rm I} +(1-\delta)\langle f_{\rm II}\rangle_{\rm II} \bigr\} \nonumber \\
&+\lambda_{\rm t}\big\{ (1+\delta)\langle |\bm{g}^{\rm I}_{\tilde{\bm{k}}}|f_{\rm I} \rangle_{\rm I}
-(1-\delta)\langle |\bm{g}^{\rm II}_{\tilde{\bm{k}}}| f_{\rm II}\rangle_{\rm II} \big\}
\big],
\label{gap-eq2_insi}
\end{align}
where $[n_{\rm c}(T)]$ indicates the integer part of $n_{\rm c}(T)=(\omega_{\rm c}/\pi T-1)/2$
with the cutoff frequency set to $\omega_{\rm c}=7\pi T_{\rm c0}$ for the numerical calculations throughout the paper.
Here, $T_{\rm c0}\equiv(2\omega_{\rm c}{\rm e}^\gamma/\pi){\rm e}^{-1/\lambda}$, where $\gamma=0.577...$ is the Euler constant,
is determined via $\frac{1}{\lambda}\equiv \sum_{n=0}^{[n_{\rm c}(T_{\rm c0})]}1/(n+1/2)$.
The quasiclassical Green's functions $g_l$ and $f_l$ ($l={\rm I,~II}$) are given in Appendix \ref{appendixA}. 
In the limit of $T\rightarrow T_{\rm c}$,
the coupling constants for the spin-singlet and triplet attraction force channels are determined from the linearized gap equation at $\bm{H}=\bm{0}$ in the clean limit through
\begin{widetext}
\begin{align}
\lambda_{\rm s}&=
\frac{
2\lambda\nu -\lambda_{\rm m}
\bigl\{
(1+\delta)\nu\langle |\bm{g}^{\rm I}_{\tilde{\bm{k}}}|\rangle_{\rm I}-(1-\delta)\nu\langle |\bm{g}^{\rm II}_{\tilde{\bm{k}}}| \rangle_{\rm II}
+(1+\delta)\langle |\bm{g}^{\rm I}_{\tilde{\bm{k}}}|^2 \rangle_{\rm I}+(1-\delta)\langle |\bm{g}^{\rm II}_{\tilde{\bm{k}}}|^2 \rangle_{\rm II}
\bigr\}
}
{
2\nu+(1+\delta)\langle |\bm{g}^{\rm I}_{\tilde{\bm{k}}}| \rangle_{\rm I} -(1-\delta)\langle |\bm{g}^{\rm II}_{\tilde{\bm{k}}}|\rangle_{\rm II}
},
\label{coupling_2band_singlet}
\\
\lambda_{\rm t}&=
\frac{
2\lambda-\lambda_{\rm m}\bigl\{ 2\nu+(1+\delta)\langle |\bm{g}^{\rm I}_{\tilde{\bm{k}}}| \rangle_{\rm I}-(1-\delta)\langle |\bm{g}^{\rm II}_{\tilde{\bm{k}}}| \rangle_{\rm II} \bigr\}
}
{
(1+\delta)\nu\langle |\bm{g}^{\rm I}_{\tilde{\bm{k}}}| \rangle_{\rm I}-(1-\delta)\nu\langle |\bm{g}^{\rm II}_{\tilde{\bm{k}}}| \rangle_{\rm II}
+(1+\delta)\langle |\bm{g}^{\rm I}_{\tilde{\bm{k}}}|^2\rangle_{\rm I}+(1-\delta)\langle |\bm{g}^{\rm II}_{\tilde{\bm{k}}}|^2\rangle_{\rm II}
},
\label{coupling_2band_triplet}
\end{align}
\end{widetext}
respectively, where $\lambda_{\rm m}$ is the coupling constant for the mixing channel, $\nu=\left. \psi_{\rm s}/d_{\rm t} \right|_{T\rightarrow T_{\rm c}-0}$ is the parity mixing ratio, and
\begin{equation}
\frac{1}{\lambda}\approx
\ln\left( T\over T_{\rm c0}\right)+\sum_{n=0}^{[n_{\rm c}(T)]}
\frac{2}{2n+1}.
\end{equation}

\section{Numerical results}
\label{sec:IV}
\subsection{Transition line}
\begin{figure*}[tb]
  \begin{center}
     \begin{tabular}{p{90mm}p{90mm}}
      \resizebox{90mm}{!}{\includegraphics{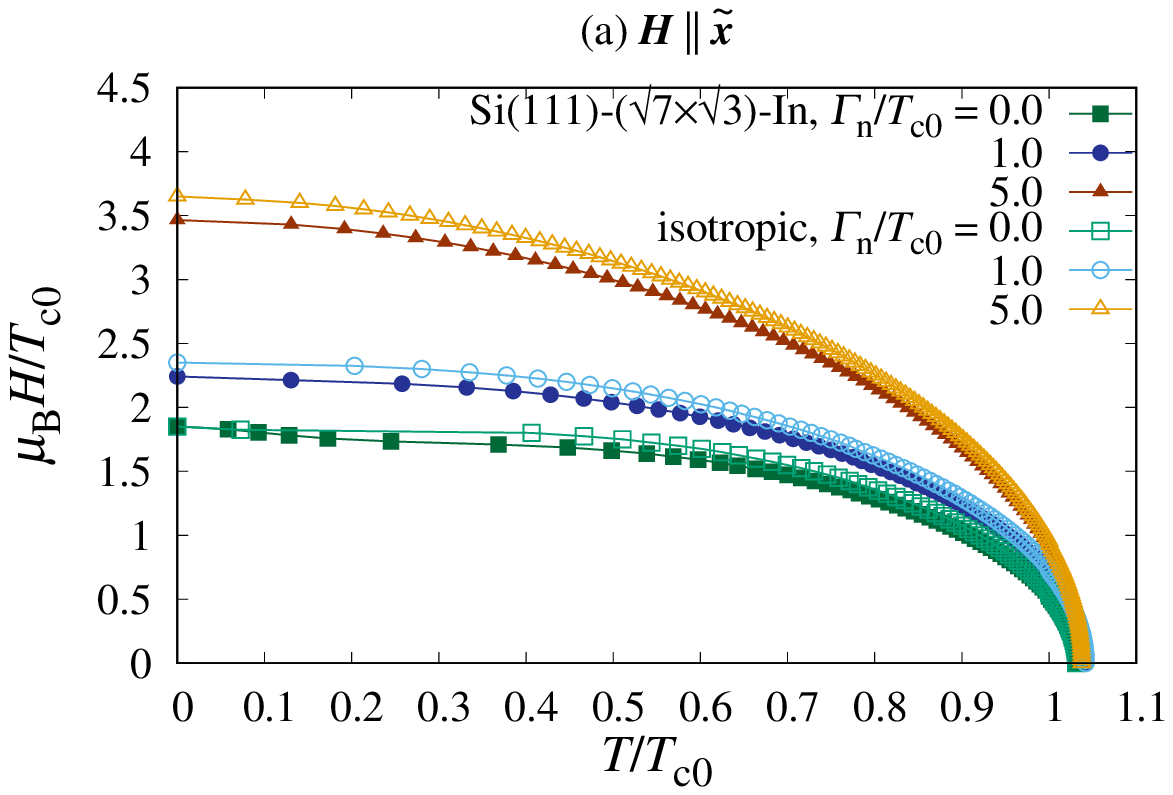}}&
      \resizebox{90mm}{!}{\includegraphics{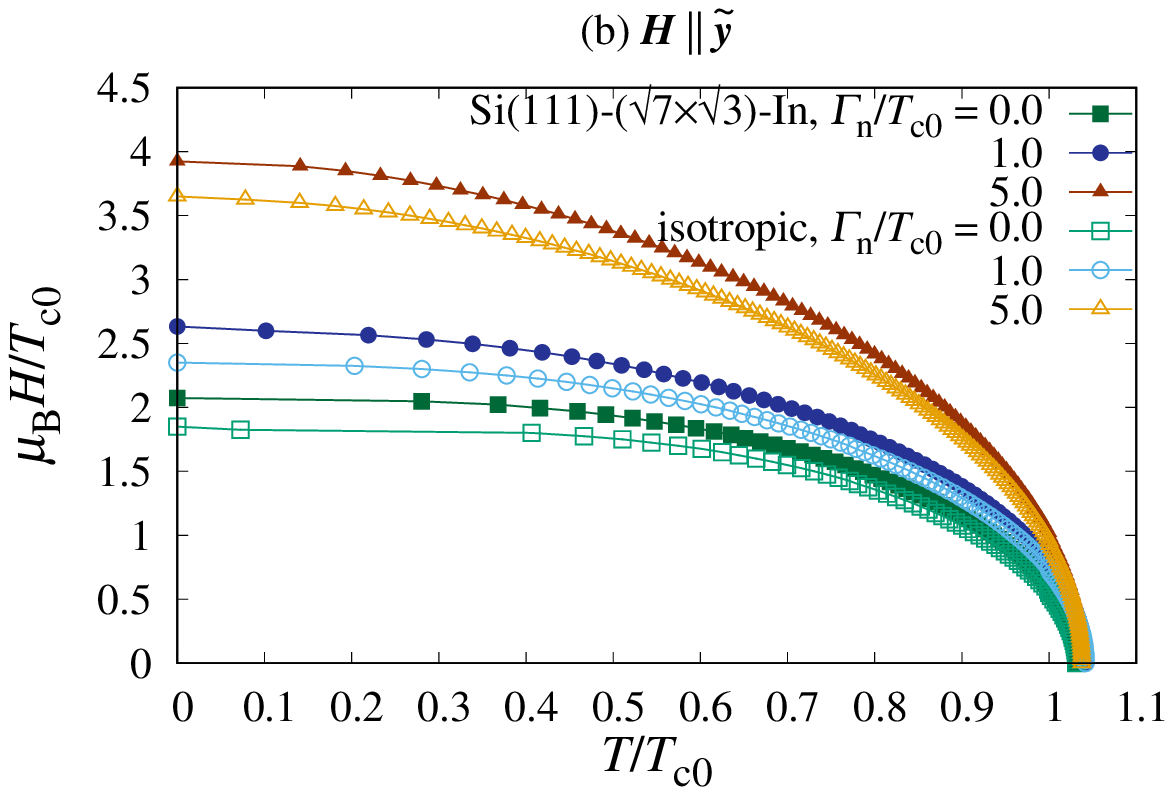}}
    \end{tabular}
\caption{
\label{fig2}
(Color online)
Temperature dependence of an in-plane critical magnetic field oriented parallel to the (a) $x$ and (b) $y$ axes for $s$-wave pairing upon varying the normal-state scattering rate $\varGamma_{\rm n}$.
The filled and open symbols represent the data for the FS of Si(111)--$(\sqrt{7}\times\sqrt{3})$-In and an isotropic system, respectively.
The difference in the DOSs between the two split FSs is set to $\delta=0$.
}
  \end{center}
\end{figure*}
\begin{figure*}[tb]
  \begin{center}
    \begin{tabular}{p{90mm}p{90mm}}
      \resizebox{90mm}{!}{\includegraphics{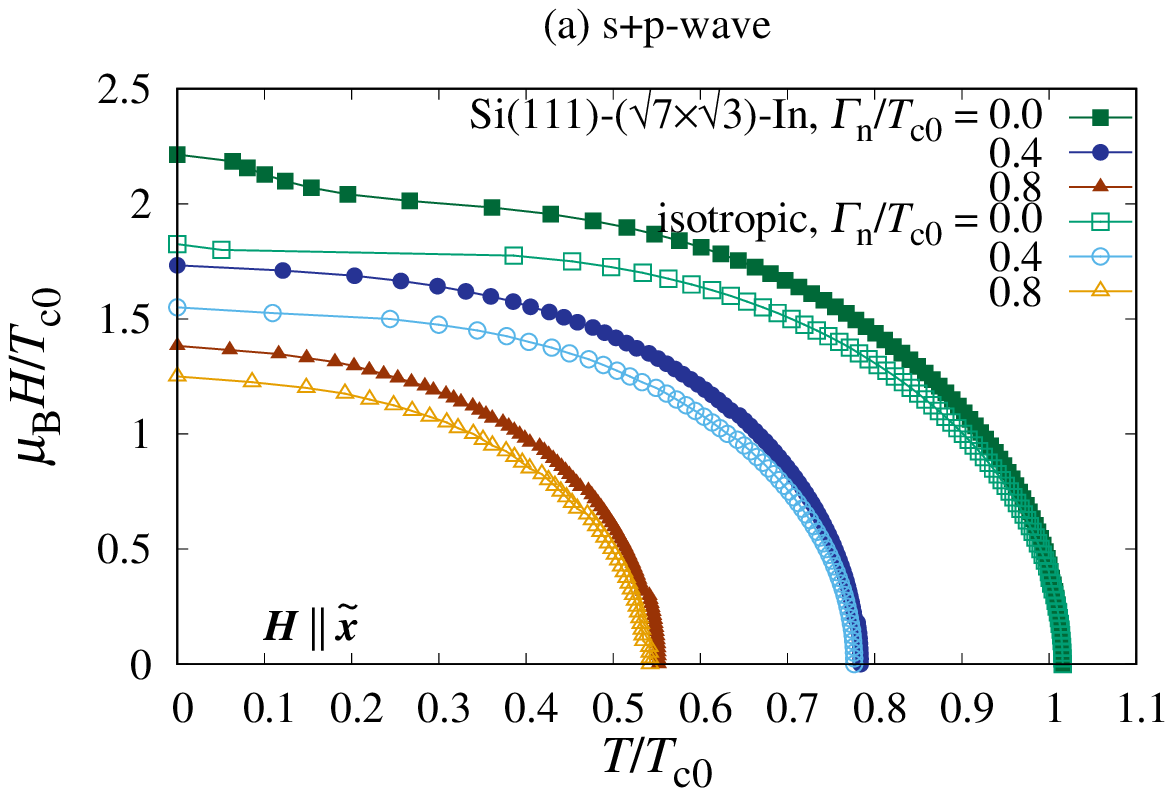}}&
      \resizebox{90mm}{!}{\includegraphics{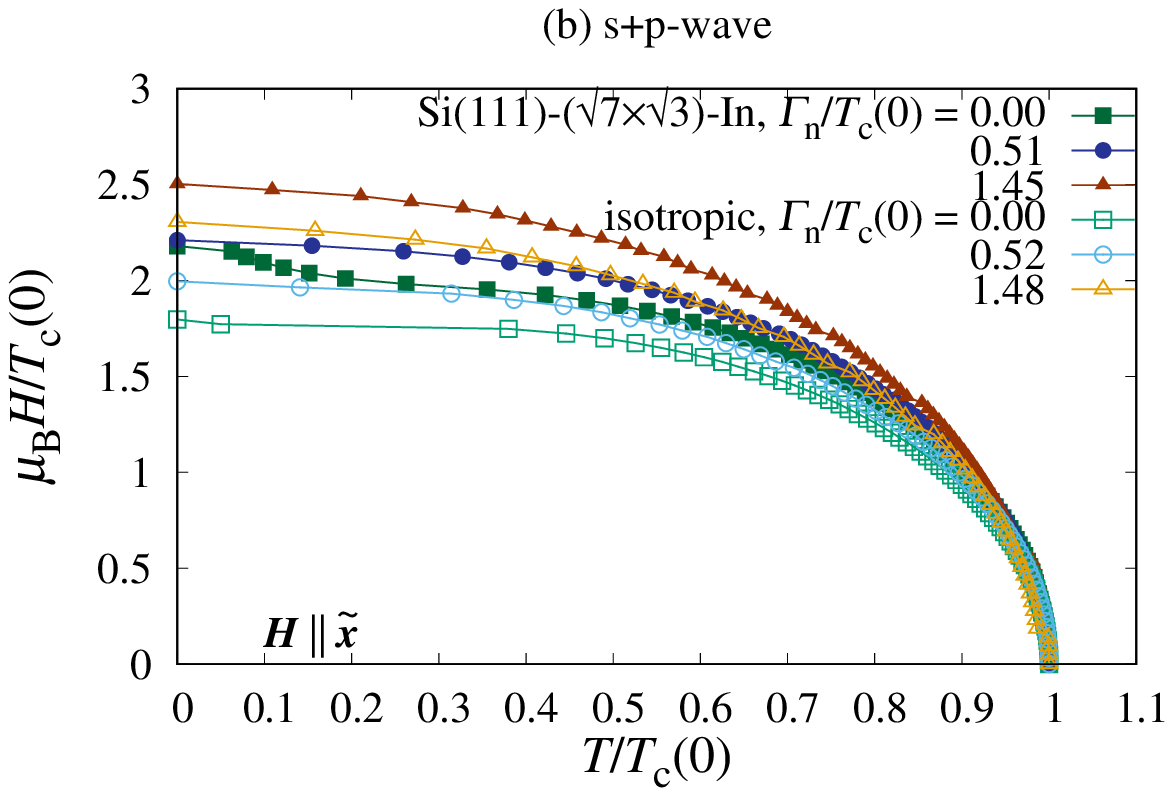}}
    \end{tabular}
\caption{
\label{fig4}
(Color online)
Temperature dependence of an in-plane critical magnetic field oriented parallel to the $x$ axis for the $s$+$p$-wave pairing upon varying the normal-state scattering rate $\varGamma_{\rm n}$.
The filled and open symbols represent the data for the FS of Si(111)--$(\sqrt{7}\times\sqrt{3})$-In and an isotropic system, respectively.
The difference in the DOSs between the two split FSs is set to $\delta=0$.
The parity mixing ratio and the coupling constant for the mixing channel are $\nu=0.5$ and $\lambda_{\rm m}=0.1$, respectively.
The temperature and magnetic field are normalized by (a) $T_{\rm c0}$ and (b) $T_{\rm c}(0)$, respectively.
}
  \end{center}
\end{figure*}
\begin{figure*}[tb]
  \begin{center}
    \begin{tabular}{p{90mm}p{90mm}}
      \resizebox{90mm}{!}{\includegraphics{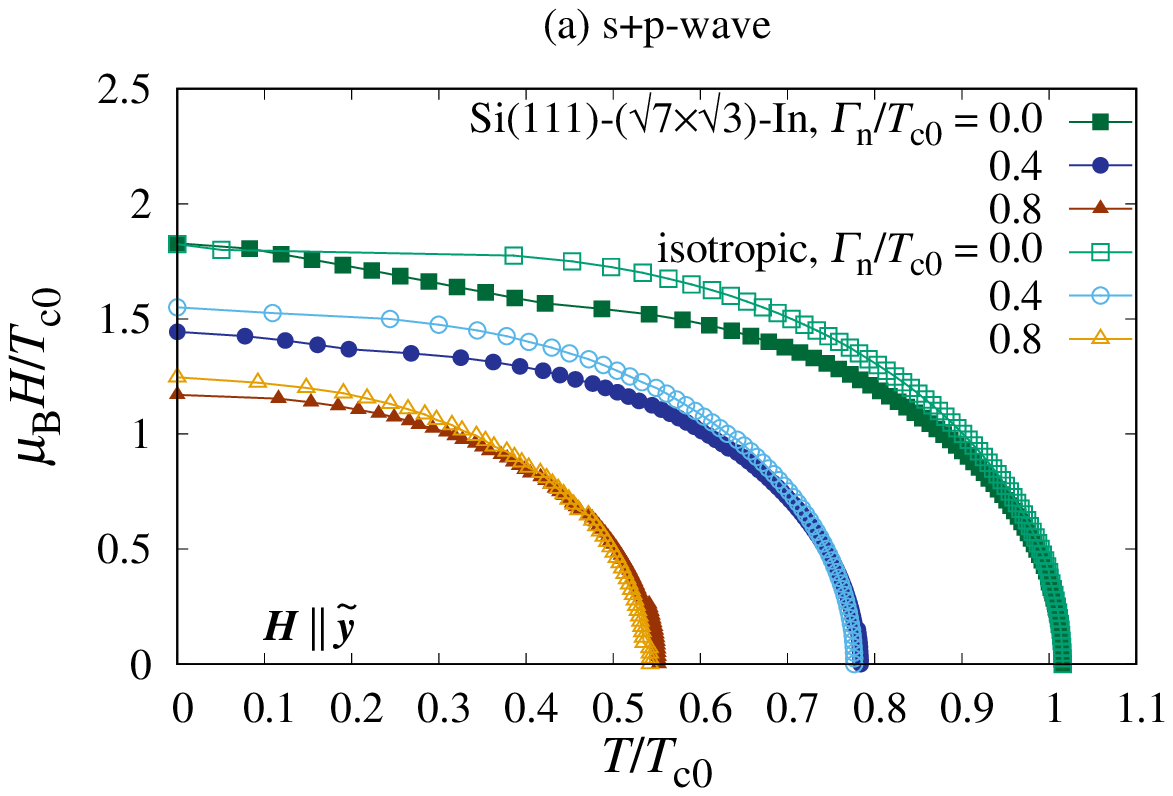}}&
      \resizebox{90mm}{!}{\includegraphics{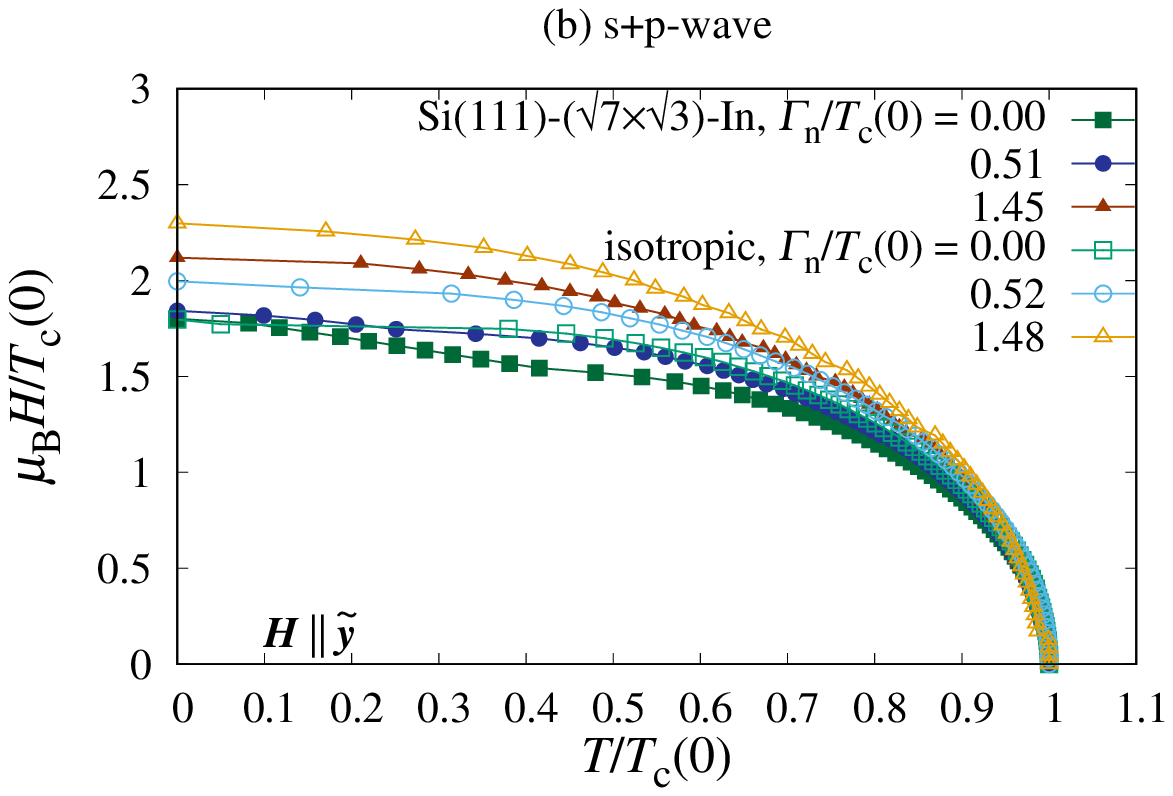}}
    \end{tabular}
\caption{
\label{fig5}
(Color online)
Temperature dependence of an in-plane critical magnetic field oriented parallel to the $y$ axis for the $s$+$p$-wave pairing upon varying the normal-state scattering rate $\varGamma_{\rm n}$.
The filled and open symbols represent the data for the FS of Si(111)--$(\sqrt{7}\times\sqrt{3})$-In and an isotropic system, respectively.
The difference in the DOSs between the two split FSs is set to $\delta=0$.
The parity mixing ratio and the coupling constant for the mixing channel are $\nu=0.5$ and $\lambda_{\rm m}=0.1$, respectively.
The temperature and magnetic field are normalized by (a) $T_{\rm c0}$ and (b) $T_{\rm c}(0)$, respectively.
}
  \end{center}
\end{figure*}

We numerically solve Eqs. (\ref{gap-eq1_isotropic}) and (\ref{gap-eq2_isotropic}) [or Eqs. (\ref{gap-eq1_insi}) and (\ref{gap-eq2_insi})]
for an isotropic [or Si(111)--($\sqrt{7}\times\sqrt{3}$)-In] FS in an iterative manner to achieve self-consistency, $\Delta Q<1\times 10^{-6}$.
Here, $\Delta Q\equiv\max_{{\rm i}\omega_n}\left[ |Q_{\rm new}({\rm i}\omega_n)-Q_{\rm old}({\rm i}\omega_n)|\right]$
with $Q({\rm i}\omega_n)$ being $\psi_{\rm s}$, $d_{\rm t}$, $\sigma_{\rm g}({\rm i}\omega_n)$, or $\sigma_{\rm f}({\rm i}\omega_n)$.
We use the bisection method for ${\rm Re}\left[\psi_{\rm s} (d_{\rm t})(T,\bm{H})\right]-\varepsilon=0$
to obtain the numerical solutions of $H^{||}_{\rm c2}(T)$ for $0.05T_{\rm c0}\le T\le 1.2T_{\rm c0}$.
The numerical solution at low temperature ($T<0.05T_{\rm c0}$) is regarded as $H^{||}_{\rm c2}(0)$.
The constant energy shift used in the bisection method is always set to $\varepsilon/T_{\rm c0}=1\times10^{-4}$.
The DOS difference between the split FSs is set to $\delta=0$ in accordance with a uniform state.
In the presence of the DOS difference, the contribution of the outer FS I originating from the band 400 with a larger DOS at $\varepsilon_{\rm F}$ becomes more prominent,
and therefore the spin structure on the outer FS I is more influential for $H^{||}_{\rm c2}$.

Figures \ref{fig2}(a) and \ref{fig2}(b) show the temperature dependence of an in-plane critical magnetic field $H^{||}_{\rm c2}(T)$ in the case of an $s$-wave pairing for $\bm{H}~||~\tilde{\bm{x}}$ and $\bm{H}~||~\tilde{\bm{y}}$, respectively.
Here, $\tilde{\bm{x}}$ ($\tilde{\bm{y}}$) denotes the unit vector in the direction of the $x$ ($y$) axis, taken to be parallel to the $k_x$ ($k_y$) direction in Fig.~\ref{fig1}.
For both the Si(111)--($\sqrt{7}\times\sqrt{3}$)-In and isotropic FSs,
$H^{||}_{\rm c2}$ is enhanced with increasing $\varGamma_{\rm n}$ irrespective of the field direction,
in accordance with the results reported by Dimitrova and Feigel'man for an $s$-wave pairing on an isotropic FS \cite{dimitrova2007}.
No change in the superconducting transition temperature at zero field $T_{\rm c}(0)$ against nonmagnetic scattering reflects Anderson's first theorem \cite{anderson1959,memo}.
For 2D superconductors such as ultrathin amorphous films,
as the film thickness is reduced,
the disorder and quantum fluctuations of the superconducting phase increase,
forming localized unpaired electrons, which lead to the suppression of $T_{\rm c}$ \cite{strongin1970,jaeger1986,haviland1989,finkelstein1994}.
Here, for the sake of simplicity, we disregarded such effects because disorder is weak in a highly crystalline atomic layer judging from the small value of the normal-state sheet resistance,
and a sharp transition to SC was observed by electron transport measurements \cite{yoshizawa2021}.
As shown in Fig. \ref{fig2}(a), for $\bm{H}~||~\tilde{\bm{x}}$, $H^{||}_{\rm c2}(T)$ was larger for the isotropic FS than for the Si(111)--($\sqrt{7}\times\sqrt{3}$)-In FS.
By contrast, as shown in Fig. \ref{fig2}(b), for $\bm{H}~||~\tilde{\bm{y}}$,
$H^{||}_{\rm c2}(T)$ was larger for the Si(111)--($\sqrt{7}\times\sqrt{3}$)-In FS than for the isotropic FS over the entire temperature range.
Thus, in the case of the anisotropic FS,  $H^{||}_{\rm c2}$ is not always enhanced.
The enhancement depends on the relation between the field direction and the spin structure on the FS.

Figures \ref{fig4}(a) and \ref{fig4}(b) present plots of $H^{||}_{\rm c2}(T)$ normalized by $T_{\rm c0}$ and $T_{\rm c}(0)$, respectively, in the case of $s$+$p$-wave pairing for $\bm{H}~||~\tilde{\bm{x}}$.
Note that $H^{||}_{\rm c2}(T)$ for $\psi_{\rm s}$ and $d_{\rm t}$ show almost the same profile, but the amplitudes of $\psi_{\rm s}$ and $d_{\rm t}$ vary
depending on the parity mixing parameters $\nu$ and $\lambda_{\rm m}$.
Here, they are set to $\nu=0.5$ ({\it i.e.}, dominant $p$-wave component $d_{\rm t}$) and $\lambda_{\rm m}=0.1$, respectively.
The variation of the transition line upon changing $\nu$ and $\lambda_{\rm m}$ are discussed in Appendix \ref{appendixC}.
As shown in Fig. \ref{fig4}(a), nonmagnetic scattering is detrimental to $T_{\rm c}$ because of the dominance of $d_{\rm t}$ over $\psi_{\rm s}$.
More specifically,
by examining the linearized gap equation in the case of parity mixing,
we notice that the scale factors of the following quantities are different as long as $\nu$ is finite:
\begin{align}
1+\frac{\sigma_{\rm g}}{\omega_n}
&=1+\frac{\varGamma_{\rm n}}{|\omega_n |}\equiv\eta(\omega_n),
\label{scalefactor-s}\\
1+\frac{\sigma_{\rm f}}{\psi_{\rm s}}
&=\eta(\omega_n)\left( 1+\frac{\delta}{\nu}\right)-\frac{\delta}{\nu}
\label{scalefactor-t}.
\end{align}
Therefore, the scale factors of Eqs. (\ref{scalefactor-s}) and (\ref{scalefactor-t}) in the anomalous Green's function do not cancel out, and thus the nonmagnetic scattering affects $T_{\rm c}$.
Fig. \ref{fig4}(b) shows that the enhancement of $\mu_{\rm B}H^{||}_{\rm c2}/T_{\rm c}(0)$ with increasing $\varGamma_{\rm n}$ remains,
but it is rather weak in the case of the dominant $p$-wave pairing compared with the pure $s$-wave pairing \cite{dimitrova2007} [see Fig. \ref{fig2}(a)].

The clean-limit data for the isotropic FS (open squares) in Fig. \ref{fig4}(b),
$\mu_{\rm B}H_{\rm c2}^{||}(T \approx 0)/T_{\rm c}(0)$, show the Pauli-limiting field for an isotropic Rashba system,
which is estimated via $\sqrt{2}H^{\rm P}\approx 1.77T_{\rm c}(0)/\mu_{\rm B}$ with
$H^{\rm P}=\sqrt{2}\psi_{\rm s}/g\mu_{\rm B}$ as the conventional Pauli-limiting field.
We use the weak-coupling Bardeen-Cooper-Schrieffer (BCS) ratio and the electronic $g$-factor, $g=2$.
Thus, $H^{||}_{\rm c2}(T \approx 0)$ clearly exceeds $\sqrt{2}H^{\rm P}$.
It turns out that the $H^{||}_{\rm c2}(T \approx 0)$ enhancement appears also in the dominant $p$-wave case.
The enhancement of $H^{||}_{\rm c2}(T \approx 0)$ is a result from both the anisotropic spin texture and disorder effect.

In the case of $\bm{H}~||~\tilde{\bm{y}}$ (Fig. \ref{fig5}),
$\mu_{\rm B} H^{||}_{\rm c2}/T_{\rm c}(0)$ is also enhanced with increasing $\varGamma_{\rm n}$,
but in the case of the anisotropic FS this enhancement is suppressed,
in contrast to the case of $\bm{H}~||~\tilde{\bm{x}}$.
The $H_{\rm c2}^{||}$ enhancement is dependent on the in-plane field direction,
demonstrating that the anisotropic spin texture does not always increase $H_{\rm c2}^{||}$.

A slight upturn of $H^{||}_{\rm c2}$ at low temperatures is observed in Figs. \ref{fig2}--\ref{fig5} for Si(111)--($\sqrt{7}\times\sqrt{3}$)-In in the clean limit.
Because this behavior is absent in the isotropic Rashba system, it is ascribed to the anisotropic spin splitting and spin texture as illustrated in Fig. \ref{fig1}.
With increasing $\varGamma_{\rm n}$, the anisotropic feature is considered to be washed out,
resulting in no upturn of $H^{||}_{\rm c2}$ as observed in the data for $\varGamma_{\rm n}\neq 0$ in Figs. \ref{fig2}--\ref{fig5}.
An upturn of $H^{||}_{\rm c2}$ in the absence of helical modulation was also reported in Ref. \cite{nakamura2013},
although it was attributed to the orbital degrees of freedom together with the Rashba SOC.
\subsection{Magnetic-field resilience}
\begin{figure}[tb]
  \begin{center}
    \begin{tabular}{p{90mm}}
      \resizebox{90mm}{!}{\includegraphics{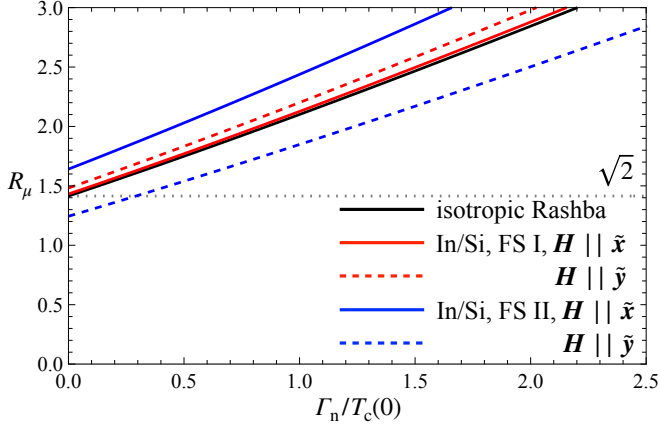}}
    \end{tabular}
\caption{
\label{fig6}
(Color online)
Dependence of the field-resilience factor $R_\mu$ on $\varGamma_{\rm n}$ for the isotropic Rashba system and Si(111)--($\sqrt{7}\times\sqrt{3}$)-In (In/Si). 
}
  \end{center}
\end{figure}
The first equation for determining $H_{\rm c2}(T)$ in superconductors with Pauli paramagnetism was derived by incorporating the SOC only through the spin--orbit scattering time \cite{maki1966,klemm1975}.
Later, in the clean limit, an equation for $H_{\rm c2}(T)$ with the textured spin structure being explicitly incorporated was obtained \cite{barzykin2002,frigeri2004,smidman2017},
thereby allowing the suppression of the paramagnetic depairing by the SOC to be discussed.
The $H^{||}_{\rm c2}(T)$ line was computed for the 2D Rashba model taking account of the contribution that is not incorporated into the impurity self-energy \cite{dimitrova2007}.
Below, we present the analytic expression for $H^{||}_{\rm c2}(T\approx T_{\rm c}(0))$ in the case of parity mixing, taking account of both the impurity scattering and suppression of paramagnetic depairing:
\begin{align}
&\frac{T_{\rm c}(0)-T_{\rm c}}{T_{\rm c}(0)}
=
\frac{\zeta(3,1/2+\gamma_{\rm n})}{4\pi^2}  \frac{ \mu^2_{\rm B} \langle   ( \bar{\bm{g}}_{\tilde{\bm{k}}}\cdot \bm{H}  )^2  \rangle_{\tilde{\bm{k}}}}{T^2_{\rm c}(0)}\nonumber\\
&+\frac{\lambda_{\rm s}-(1+\delta)[ \lambda_{\rm s}(1+\langle |\bm{g}_{\tilde{\bm{k}}}| \rangle/\nu)+\lambda_{\rm m}(\langle|\bm{g}_{\tilde{\bm{k}}}| \rangle+1/\nu)]}
{\lambda_{\rm s}(1+\delta)(\lambda_{\rm s}+\lambda_{\rm m})(1+1/\nu)},
\label{hc2-line}
\end{align}
where $\gamma_{\rm n}\equiv\varGamma_{\rm n}/2\pi T_{\rm c}(0)$ and $\zeta(z,a)=\sum_{n=0}^{\infty}1/(a+n)^z$ ($z\in \mathbb{C}$, $a$: constant) is the Hurwitz zeta function.
Note that Eq.~(\ref{hc2-line}) is derived by assuming that the FS is infinitesimally split and $|\bar{\bm{g}}_{\tilde{\bm{k}}}\cdot \mu_{\rm B}\bm{H}/\pi T_{\rm c}|\ll1$.
In Eq.~(\ref{hc2-line}), $T_{\rm c}(0)$ should be read as the zero-field transition temperature without the parity mixing and the DOS difference.
For generic cases characterized by the antisymmetric orbital vector $\bm{g}_{\tilde{\bm{k}}}$,
the effective field is $H_{\rm eff} \equiv H_\mu/R_\mu$ with
\begin{equation}
R_\mu \equiv  \left[\frac{\zeta(3,1/2+\gamma_{\rm n})}{7\zeta(3)}\langle g^2_{\mu} (\tilde{\bm{k}}) \rangle_{\tilde{\bm{k}} }\right]^{-1/2}
\label{field-resilience}
\end{equation}
for $\bm{H}~||~\tilde{\bm{\mu}}$.
Consequently, the Pauli limiting field ($||~\tilde{\bm{\mu}}$) is determined via $H_{\rm eff}=H^{\rm P}$ as
\begin{equation}
H^{\rm P}_\mu=R_\mu H^{\rm P}.
\end{equation}
The physical meaning of $R_\mu$ is interpreted as the {\it magnetic-field resilience} of SC for $H_\mu$.
The $H^{||}_{\rm c2}$ enhancement from $\sqrt{2}H^{\rm P}$ is roughly judged from the condition that
$\langle g^2_\mu(\tilde{\bm{k}})\rangle_{\tilde{\bm{k}}} <1/2$.
The dependence of $R_\mu$ on $\varGamma_{\rm n}$ is plotted in Fig.~\ref{fig6}.
When evaluating Eq.~(\ref{field-resilience}),
we phenomenologically replace the average on the infinitesimally split FS $\langle \cdots\rangle_{\tilde{\bm{k}}}$
with that on the significantly split FSs I and II $\langle \cdots\rangle_{\rm I,II}$
to apply $R_\mu$ to the FS of Si(111)--($\sqrt{7}\times\sqrt{3}$)-In (In/Si). 
$R_\mu$ increases monotonically with respect to $\varGamma_{\rm n}$, in accordance with the $H^{||}_{\rm c2}$ enhancement with increasing $\varGamma_{\rm n}$.
For $\bm{H}~||~\tilde{\bm{x}}$, the $R_x$ values for both of the In/Si FSs I and II are larger than $R_\mu$ for the isotropic Rashba system.
However, in the case of $\bm{H}~||~\tilde{\bm{y}}$, the magnitude of $R_y$ for In/Si relative to that for the isotopic Rashba system
depends on the details of the FS and the spin texture,
as reflected in $\langle g^2_y(\tilde{\bm{k}})\rangle_{\rm I}=0.455<1/2$ and $\langle g^2_y(\tilde{\bm{k}})\rangle_{\rm II}=0.646>1/2$. 

In the absence of parity mixing ($\lambda_{\rm m}=0$ and $\nu\rightarrow \infty$) and the DOS difference between the split FSs ($\delta=0$),
Eq.~(\ref{hc2-line}) for an isotropic 2D Rashba superconductor [$\bar{\bm{g}}_{\tilde{\bm{k}}}=(-\sin \phi_k,\cos \phi_k,0)$]
in the clean limit ($\gamma_{\rm n}=0$)
reduces to the result reported by Barzykin and Gor'kov \cite{barzykin2002}:
\begin{equation}
\frac{T_{\rm c}(0)-T_{\rm c}}{T_{\rm c}(0)}=\frac{7\zeta(3)}{4\pi^2}    \frac{\mu^2_{\rm B}  (H/\sqrt{2})^2}{T^2_{\rm c}(0)}.
\end{equation}
Here, $\sqrt{\langle(\bar{\bm{g}}_{\tilde{\bm{k}}}\cdot\bm{H})^2\rangle_{\tilde{\bm{k}}}}=H/\sqrt{2}$ can be viewed as an effective field for the isotropic Rashba SC.
Thus, the enhancement of the Pauli limiting field is limited to $\sqrt{2}H^{\rm P}$.


\subsection{Comparison with experimental $H^{||}_{\rm c2}(T)$ data}
\begin{figure}[tb]
  \begin{center}
    \begin{tabular}{p{90mm}}
      \resizebox{90mm}{!}{\includegraphics{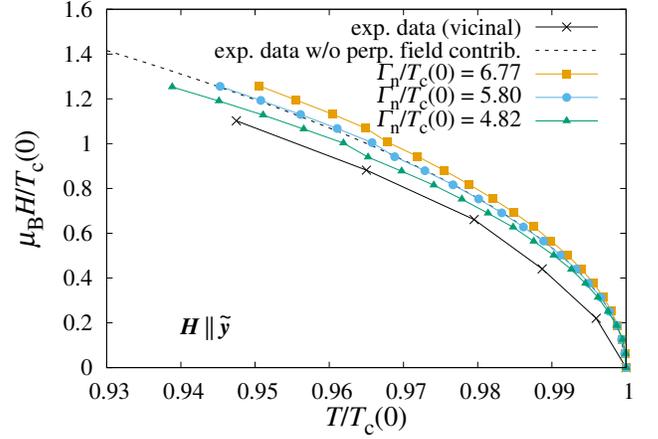}}
    \end{tabular}
\caption{
\label{fig7}
(Color online)
Comparison of the theoretically evaluated in-plane critical magnetic field for the $s$-wave pairing
with the experimental data for the $(\sqrt{7}\times\sqrt{3})$-In sample on a vicinal Si(111) surface.
The magnetic field is parallel to the $y$ axis.
The difference in the DOSs between the two split FSs is $\delta=0.0589$.
}
  \end{center}
\end{figure}
In Fig. \ref{fig7}, the experimental $H^{||}_{\rm c2}(T)$ data for Si(111)--($\sqrt{7}\times\sqrt{3}$)-In are compared with the numerical data for an $s$-wave pairing.
The field direction is set to $\bm{H}~||~\tilde{\bm{y}}$ in accordance with the experimental setup.
For the pair of split FSs with DOSs an order of magnitude larger than the others [see Fig.~\ref{fig1}(b)],
the DOS difference is evaluated as $\delta=0.0589$ by DFT calculations.
By subtracting a perpendicular field component [$H^{\perp}_{\rm c2}(T)\propto T_{\rm c}(0)-T$] from the experimental $H^{||}_{\rm c2}(T\approx T_{\rm c}(0))$ data,
we may compare the numerical calculation results with the contribution from the in-plane field responsible for paramagnetic depairing in the experimental data.
The experimental data of the parallel field contribution (dashed line in Fig.~\ref{fig7}) are in good agreement with the numerical data for $\varGamma_{\rm n}/T_{\rm c}(0)\approx 5.8$ in the case of an $s$-wave pairing.
Assuming the parabolic dependence $T/T_{\rm c}(0)=1-c[\mu_{\rm B}H/T_{\rm c}(0)]^2$,
we identify the quadratic coefficients $c_{\rm exp}$ and $c_{\rm num}$ for the experimental and numerical $H^{||}_{\rm c2}(T)$ data, respectively.
Regarding $c_{\rm num}$ for each $\varGamma_{\rm n}$, we obtain $c_{\rm num}(\varGamma_{\rm n})$ by polynomial fitting.
By solving $c_{\rm num}(\varGamma_{\rm n})=c_{\rm exp}$,
we obtain the more precise value of $\varGamma_{\rm n}/T_{\rm c}(0)\approx 5.9$, which is consistent with the previous study \cite{dimitrova2007}.
From the low-field data of $H^{||}_{\rm c2}(T)$, one can quantitatively estimate the normal-state scattering rate.

On the basis of Eq.~(\ref{hc2-line}), in the absence of parity mixing ($\nu\rightarrow \infty$ and $\lambda_{\rm m}=0$) and the DOS difference ($\delta=0$),
the quadratic coefficient for an isotropic Rashba system is obtained as
\begin{equation}
c(\varGamma_{\rm n})=\frac{ \zeta(3,1/2+\gamma_{\rm n})}{8\pi^2}.
\end{equation}
By solving $c(\varGamma_{\rm n})=c_{\rm exp}$, we obtain $\varGamma_{\rm n}/T_{\rm c}(0)\approx1.5$.
Thus, Eq.~(\ref{hc2-line}) somewhat underestimates $\varGamma_{\rm n}$ compared with the numerically estimated value.
This is because Eq.~(\ref{hc2-line}) is derived by assuming infinitesimally split FSs,
and thus strictly speaking it is not directly applicable to a system in the large ASOC regime.

\section{Discussion}
\label{sec:V}
In the evaluation of $H^{||}_{\rm c2}$ with the developed quasiclassical theory in the strong ASOC limit, we assume some additional approximations that ignore the following phenomena;
(a) phase modulation in the helical state, (b) localized unpaired electrons due to disorders and quantum fluctuations in 2D systems, (c) scattering off atomic steps and inter-atomic-terrace orbital motion.
In the following, we first discuss (c) and lastly give a physical picture of how the spin texture on the FS and the disorder support the field resilient superconductivity.

\subsection{Inter-atomic-terrace orbital effect}
The SC of Si(111)--($\sqrt{7}\times\sqrt{3}$)-In can be affected by possible orbital effects in an in-plane field in the case of high atomic step density.
For a vicinal substrate, the atomic step density may be as high as approximately $20/500~{\rm nm}$ \cite{yoshizawa2021}. 
Thus, the total atomic step height may reach several micrometers ($\sim100 \xi$) over the sample size
by assuming the same height of approximately $0.3$ nm \cite{yoshizawa2015} for all of the atomic steps.
For a vicinal substrate, we herein estimate the effective coherence length as $\xi\sim 40$ nm via $1/\xi=1/\xi_{\rm BCS}+1/v_{\rm F}\tau_{\rm el}$
using the following adopted parameters:
$\varDelta=0.39$ meV \cite{yoshizawa2014},
$v_{\rm F}=1.43 \times 10^6$ m/s (see Sec. \ref{sec:II}),
and $\tau_{\rm el}\sim 30$ fs \cite{yoshizawa2021},
with $\xi_{\rm BCS}= v_{\rm F}/\pi \varDelta$ denoting the BCS coherence length and $\tau_{\rm el}$ representing the elastic scattering time.
Under these circumstances, the orbital motion of electrons may be allowed owing to the sufficient thickness of the atomic terraces.
An indium atomic bilayer itself is considered not to be responsible for the orbital motion under an in-plane field.
How an in-plane critical field depends on {\it inter-atomic-terrace orbital effects} currently remains unresolved.

In the theoretical model, we assume a Si(111) substrate with a low atomic step density, that is, an atomically flat plane on a terrace that is as large as $\xi$.
We also assume that the field direction is parallel to the long dimension of the atomic terraces to avoid possible interatomic terrace orbital effects,
although in reality there are several thousand atomic steps over the sample size even when the field is parallel to the long dimension of the atomic terraces
because of the inevitable error in the cutout angle of a Si(111) substrate. 
In this study, for simplicity, we neglect the inter-atomic-terrace orbital effect in the theoretical model to show the enhancement of $H^{||}_{\rm c2}$,
suggesting that, from a theoretical perspective, the electron scattering off atomic steps inherent to a Si(111) substrate surface
is not primarily responsible for the $H^{||}_{\rm c2}$ enhancement.
The scattering from disorder within the flat atomic terrace plane may be more important for $H^{||}_{\rm c2}$ enhancement.

\subsection{Scattering rate via fitting analysis and electron transport}
For a flat Si(111) substrate,
there are three crystallographic orientations in the $\sqrt{7}\times\sqrt{3}$ structures \cite{yoshizawa2021},
while for a vicinal substrate cut from bulk Si(111) with a mis angle from a certain direction,
the crystallographic orientation of the ($\sqrt{7}\times\sqrt{3}$)-In is aligned in the same direction.
The atomic step density is higher on the vicinal surface than on the flat surface,
meaning that the density of electron scatterers is also higher.
Indeed, the rate of elastic electron scattering events in the normal state, $\varGamma_{\rm n}$, for ($\sqrt{7}\times\sqrt{3}$)-In samples on the vicinal substrate exhibits larger values.
From the inelastic scattering time evaluated by electron transport measurements for three different samples \cite{yoshizawa2021},
$\varGamma_{\rm n}$ falls in the range of $\varGamma_{\rm n}/T_{\rm c}(0)=17-27$ for the flat substrate.
By contrast, for the vicinal substrate, $\varGamma_{\rm n}/T_{\rm c}(0)=32-45$.
The values of $\varGamma_{\rm n}$ estimated via the normal-state sheet resistance
are an order of magnitude larger than the value of $\varGamma_{\rm n}/T_{\rm c}(0)\approx 5.9$
obtained from the fitting analysis of the low-field $H^{||}_{\rm c2}$ data for the $(\sqrt{7}\times\sqrt{3})$-In sample on a vicinal Si(111) surface.

Electron transport measurements over the sample size include the contribution from the scattering off atomic steps on the Si(111) surface.
However, the theoretical model does not explicitly incorporate atomic steps as a scattering source, and instead disordered regions are assumed to be weak scatterers and randomly distributed in the system.
Thus, the value of $\varGamma_{\rm n}$ evaluated via the electron transport measurements may be overestimated as an indicator of the effects of disorder on the SC of atomic-layer crystalline ($\sqrt{7}\times\sqrt{3}$)-In.
Within atomically flat terraces, highly crystalline $(\sqrt{7}\times\sqrt{3})$-In may be effectively much cleaner than observed from the electron transport.
Indeed, $H^{||}_{\rm c2}$ in the case of the vicinal substrate is not substantially different from the value for a flat substrate,
although the $H^{||}_{\rm c2}$ enhancement is expected for the vicinal substrate
because of the increased number of scattering events
due to the higher atomic step density.
This fact possibly supports the theoretical consideration regarding the effectively small $\varGamma_{\rm n}$ within atomic terraces.
In the present analysis, we adopted Born-type scatterers to model weak disorder within atomic terraces.
At the atomic steps, we speculate that the electron scattering is not simple forward scattering in the Born limit.
Instead, scattering with an arbitrary scattering phase including backward scattering in the unitary limit may occur.
The influence of such a scattering process on $H^{||}_{\rm c2}$ remains unknown.
 
\subsection{Physical picture of field resilient superconductivity}
We discuss the physics of the field resilient SC.
For simplicity, we consider an $s$-wave paring on each split isotropic FS without the DOS difference.
In the limit of large SOC, the split isotropic FSs are regarded as equivalent, and therefore the average on each split FS can be merged.
The impurity self energies are evaluated via the Green's functions at the zero field in the clean limit as $\sigma_{\rm g}\approx \varGamma_{\rm n}$ ($\omega_n>0$) and $\sigma_{\rm f}\approx (\varGamma_{\rm n}/\omega_n)\psi_{\rm s}$. 
Keeping $\sigma_{\rm f}$ in the numerator of the anomalous Green's function $f_{\rm I,II}$,
the linearized gap equation reads
\begin{equation}
\frac{1}{\lambda_{\rm s}}=2 \pi T_{\rm c}\sum_{\omega_n>0}\left\langle\frac{1}{\omega_n}\frac{1}{1+\left[\bar{\bm{g}}_{\tilde{\bm k}}\cdot \mu_{\rm B}\bm{H}/(\omega_n+\varGamma_{\rm n})\right]^2}\right\rangle_{\tilde{\bm k}}
\end{equation}
We observe the complete suppression of the paramagnetic depairing for $\bar{\bm{g}}_{\tilde{\bm k}}\cdot \mu_{\rm B}\bm{H}=0$,
but even when $\bar{\bm{g}}_{\tilde{\bm k}}\cdot \mu_{\rm B}\bm{H}\ne 0$,
with increasing $\varGamma_{\rm n}$ the influence of the magnetic field effectively gets smaller to suppress the paramagnetic depairing.
As described in Refs. \cite{michaeli2012,nam2016,yoshizawa2021}, this can be interpreted as follows. 
Due to intra-band nonmagnetic impurity scattering, the spin quantization axis of electrons traveling in the $\bm{k}$ direction is not fixed in the $\bar{\bm{g}}_{\bm k}$ direction,
but it is forced to rotate in the $\bar{\bm{g}}_{\bm{k}^\prime}$ direction to partially escape from the paramagnetic depairing.
In the Rashba SC with the large ASOC under an in-plane field, the intra-band states ($\bm{k},\hat{\bm{\sigma}}\cdot \bar{\bm{g}}_{\bm k}$) and ($-\bm{k},\hat{\bm{\sigma}}\cdot \bar{\bm{g}}_{-\bm k}$) pair up with a finite energy difference to form the zero center-of-mass momentum superconducting state.
In the energy domain, due to the impurity scattering the energy bands near the Fermi energy have an energy broadening,
allowing the states with smaller energy differences to be paired up with the zero center-of-mass momentum to higher magnetic fields.
In this way, the Rashba SC acquires the field resilience.

\section{Summary}
\label{sec:VI}
To study the SC in highly crystalline atomic-layer materials,
we formulated the quasiclassical theory of SC in the large ASOC regime with the incorporation of parity mixing, FS anisotropy, and spin texture.
We applied the developed theory to the atomic-layer crystalline material Si(111)--($\sqrt{7}\times\sqrt{3}$)-In
to calculate the in-plane critical magnetic field $H^{||}_{\rm c2}$ upon varying the normal-state scattering rate $\varGamma_{\rm n}$.
For Si(111)--($\sqrt{7}\times\sqrt{3}$)-In, we proceeded with the typical scenario of possible $H^{||}_{\rm c2}$ enhancement.
In accordance with the previous study \cite{dimitrova2007}, when $\varGamma_{\rm n}$ is increased, $H^{||}_{\rm c2}$ was enhanced in combination with the ASOC.
We found that this trend holds also in the case of parity mixing.
Furthermore, we demonstrated that the $H^{||}_{\rm c2}$ enhancement is dependent on the field direction,
meaning that the anisotropic FS and spin texture does not always enhance $H^{||}_{\rm c2}$.
To quantify the $H^{||}_{\rm c2}$ enhancement relative to the Pauli limiting field for an isotropic Rashba SC,
we proposed the {\it magnetic-field resilience} of SC,
which incorporates impurity scattering and details of the FS and spin texture.
Next, we extracted the value of $\varGamma_{\rm n}$
by numerically and analytically fitting the experimental $H^{||}_{\rm c2}$ data for a ($\sqrt{7}\times\sqrt{3}$)-In sample on a vicinal Si(111) surface. 
Finally, the possible inter-atomic-terrace orbital effect and normal-state electron scattering were discussed focusing on the role of atomic steps.

\section*{Acknowledgments}
We thank S. Ichinokura for discussions in the early stage of the research. 
The numerical calculations were performed on XC40 and Yukawa-21 at the Yukawa Institute for Theoretical Physics, Kyoto University. 
This work was supported by JSPS KAKENHI (Grant Nos. JP18H01876, JP21H01817, JP19H05823, JP20K05314, and JP22H01961).

\appendix
\counterwithin{figure}{section}

\section{Quasiclassical theory in the strong spin-orbit coupling limit}
\label{appendixA}
In the band basis representation, only the Zeeman field should be viewed as a perturbation,
as opposed to previous studies  on Rashba \cite{higashi2014imp} (resp. multilayered Rashba \cite{higashi2014,higashi2016}) superconductors, where the ASOC and the impurity self-energy (resp. the ASOC and Zeeman field) were treated as perturbations.
One may integrate the $4 \times 4$ Green's function in $2\times2$ Nambu space and $2\times2$ spin space $\check{G}_l({\rm i}\omega_n,\bm{k})=\check{G}_l({\rm i}\omega_n,\tilde{\bm k},\xi^l_{\bm k})$
with respect to $\xi^l_{\bm k}$ instead of $\xi_{\bm k}$ for significantly split FSs I and II, respectively:
\begin{align}
\check{g}_l({\rm i}\omega_n,\tilde{\bm k})
&\equiv
\check{\tau}_3\oint{\rm d}\xi^l_{\bm k}
\left(
\begin{array}{cc}
\hat{G}_l & \hat{F}_l  \\
\hat{\bar{F}}_l& \hat{\bar{G}}_l
\end{array}
\right)_{{\rm i}\omega_n,{\bm{k}}}\nonumber \\
&\equiv-{\rm i}\pi
 \left(
\begin{array}{cc}
\hat{g}_l & {\rm i}\hat{f}_l  \\
-{\rm i}\hat{\bar{f}}_l& -\hat{\bar{g}}_l
\end{array}
\right)_{{\rm i}\omega_n,\tilde{\bm{k}}},
\end{align}
where $\tilde{\bm k}$ is the direction of the relative momentum and $\omega_n=(2n+1)\pi T$ is the Matsubara frequency for fermions.
$\oint{\rm d}\xi^l_{\bm k}\cdots$ indicates that the contributions from poles close to $\varepsilon_{\rm F}$ for each split FS are taken into account.

The Eilenberger equation for spatially uniform systems, where $\bm{\nabla}\check{g}_l=\check{\bm{0}}$,
in the strong ASOC regime is
\begin{align}
&\left[ {\rm i}\tilde{\omega}_n\check{\tau}_3-\check{\tilde{\varDelta}}_l({\rm i}\omega_n,\tilde{\bm{k}})\pm\mu_{\rm B}\check{\bar{\bm{g}}}^l_{\tilde{\bm{k}}}\cdot \bm{H},\check{g}_l(i\omega_n,\tilde{\bm{k}})\right]=\check{0},
\label{eilenberger-eq}\\
&\check{\tilde{\varDelta}}_{l}({\rm i}\omega_n,\tilde{\bm{k}})=
 \left(
\begin{array}{cc}
0 & \tilde{\varDelta}_l({\rm i}\omega_n,\tilde{\bm{k}}) \\
 -\tilde{\varDelta}^\ast_l({\rm i}\omega_n,\tilde{\bm{k}})& 0
\end{array}
\right)\nonumber \\
&\equiv
 \left(
\begin{array}{cc}
0 & \varDelta_l(\tilde{\bm{k}})+\sigma_{\rm f}({\rm i}\omega_n) \\
-(\varDelta^\ast_l(\tilde{\bm{k}})+\bar{\sigma}_{\rm f}({\rm i}\omega_n))& 0
\end{array}
\right),
\end{align}
where $\tilde{\omega}_n\equiv \omega_n+\sigma_{\rm g}({\rm i}\omega_n)$, $\varDelta_l(\tilde{\bm{k}})=\psi_{\rm s}\pm d_{\rm t}|\bm{g}^l_{\tilde{\bm{k}}}|$,
and $\check{\bar{\bm{g}}}^l_{\tilde{\bm{k}}}={\rm diag}(\bar{\bm{g}}^l_{\tilde{\bm{k}}},-\bar{\bm{g}}^l_{\tilde{\bm{k}}} )$ with $\bar{\bm{g}}^l_{\tilde{\bm{k}}}\equiv \bm{g}^l_{\tilde{\bm{k}}}/|\bm{g}^l_{\tilde{\bm{k}}}|$, and 
$+$ and $-$ correspond to $l=$ I and II, respectively. 
Assuming $s$-wave scattering in the Born limit,
the nonmagnetic impurity scattering is incorporated through the self-energy [see Eqs. (\ref{sigma_g}) and (\ref{sigma_f})]:
\begin{align}
\sigma_{\rm g}({\rm i}\omega_n)&=\frac{\varGamma_{\rm n}}{2}\left[(1+\delta)\langle g_{\rm I}({\rm i}\omega_n,\tilde{\bm k})\rangle_{\rm I}+(1-\delta)\langle g_{\rm II}({\rm i}\omega_n,\tilde{\bm k})\rangle_{\rm II}\right],\\
\sigma_{\rm f}({\rm i}\omega_n)&=\frac{\varGamma_{\rm n}}{2}\left[ (1+\delta)\langle f_{\rm I}({\rm i}\omega_n,\tilde{\bm k})\rangle_{\rm I}+(1-\delta)\langle f_{\rm II}({\rm i}\omega_n,\tilde{\bm k})\rangle_{\rm II}\right],\nonumber \\
\end{align}
with $\langle \cdots\rangle_l$ indicating an average over the FS $l$.
Here, we used $g_l+\bar{g}_l=0$, which holds for a spatially uniform system \cite{hayashi2006}.
The solutions of the Eilenberger equation [Eq. (\ref{eilenberger-eq})] are readily obtained with the aid of the normalization condition $g^2_l+f_l\bar{f}_l=1$ as
\begin{align}
g_l({\rm i}\omega_n,\tilde{\bm{k}})&=\pm(\tilde{\omega}_n\mp {\rm i}\mu_{\rm B}\bar{\bm{g}}^l_{\tilde{\bm{k}}}\cdot \bm{H})/\varLambda_l({\rm i}\omega_n,\tilde{\bm{k}}),\\
f_l({\rm i}\omega_n,\tilde{\bm{k}})&=\pm \tilde{\varDelta}_l(\tilde{\bm{k}})/\varLambda_l({\rm i}\omega_n,\tilde{\bm{k}}),
\end{align}
with $\varLambda_l({\rm i}\omega_n,\tilde{\bm{k}})\equiv \sqrt{(\tilde{\omega}_n \mp {\rm i} \mu_{\rm B}\bar{\bm{g}}^l_{\tilde{\bm{k}}}\cdot  \bm{H} )^2+|\tilde{\varDelta}_l(\tilde{\bm{k}})|^2}$.
To satisfy the non-negativity of the real part of the retarded Green's function, which is directly related to the DOS,
\begin{equation}
{\rm Re}[g_l({\rm i}\omega_n\rightarrow \mp \mu_{\rm B}\bar{g}^l_{\bar{\bm k}}\cdot \bm{H}+{\rm i}\eta)]\ge 0,
\end{equation}
we note that the sign in front of the Green's function for a uniform system can be set to coincide with ${\rm sgn}[{\rm Re}(\tilde{\omega}_n)]$ (see also \cite{hayashi2013}).
Here, $\eta>0$ is used for energy smearing.

\section{Nonmagnetic scattering in spatially uniform noncentrosymmetric superconductors}
\subsection{Eilenberger equation}
The quasiclassical Green's function
\begin{align}
\check{g}(\bm{r},\tilde{\bm{k}},{\rm i}\omega_n)=-{\rm i}\pi
\left(
\begin{array}{cc}
\hat{g} & {\rm i}\hat{f} \\
 -{\rm i}\hat{\bar{f}}& -\hat{\bar{g}}
\end{array}
\right)
\end{align}
follows the Eilenberger equation.
In the case of the isotropic Rashba-type ASOC in the 2D system [$\bm{g}_{\tilde{\bm{k}}}=|\bm{g}_{\tilde{\bm{k}}}|(-\sin \phi,\cos \phi)$], the equation in the presence of disorder is
\begin{align}
&{\rm i}\bm{v}_{\rm F}(\tilde{\bm{k}})\cdot\bm{\nabla}\check{g}(\bm{r},\tilde{\bm{k}},{\rm i}\omega_n)\nonumber \\
&+\left[ {\rm i}\omega_n\check{\tau}_3-\check{\varDelta}-\check{\varSigma}-\alpha\check{\bm{g}}_{\tilde{\bm{k}}}\cdot\check{\bm{S}},\check{g}(\bm{r},\tilde{\bm{k}},{\rm i}\omega_n) \right] = \check{0},
\end{align}
with
$
\check{\tau}_3={\rm diag}(\hat{\sigma}_0,-\hat{\sigma}_0),
\check{\bm{g}}_{\tilde{\bm{k}}}={\rm diag}(\bm{g}_{\tilde{\bm{k}}}\hat{\sigma}_0,\bm{g}_{-\tilde{\bm{k}}}\hat{\sigma}_0),
\check{\bm{S}}={\rm diag}(\hat{\bm{\sigma}},\hat{\bm{\sigma}}^{\top}),
\hat{\bm{\sigma}}^{\top}=-\hat{\sigma}_y\hat{\bm{\sigma}}\hat{\sigma}_y$,
\begin{align}
\check{\varDelta}&=
\left(
\begin{array}{cc}
\hat{0} & \hat{\varDelta}(\bm{r},\tilde{\bm{k}}) \\
 -\hat{\varDelta}^\dag(\bm{r},\tilde{\bm{k}})& \hat{0} 
\end{array}
\right),\\
\hat{\varDelta}(\bm{r},\tilde{\bm{k}})&=
\left[ \psi_{\rm s}(\bm{r})\hat{\sigma}_0 + \bm{d}_{\tilde{\bm{k}}}(\bm{r}) \cdot \hat{\bm{\sigma}}  \right]{\rm i}\hat{\sigma}_y,\\
\bm{d}_{\tilde{\bm{k}}}(\bm{r})&=d_{\rm t}(\bm{r})\bm{g}_{\tilde{\bm{k}}}
\end{align}
as the order parameter,
and $\check{\varSigma}({\rm i}\omega_n,\bm{r},\tilde{\bm{k}})\equiv \{ \hat{\varSigma}_{ij} \}_{i,j=1,2}$ as the impurity self-energy.
The Eilenberger equation with respect to each component in Nambu space is recast as
\begin{subequations}
\begin{align}
\partial \hat{g}_0&+{\rm i}\alpha \bm{g}_k \cdot (\hat{\bm{\sigma}}\hat{g}_0-\hat{g}_0\hat{\bm{\sigma}})
+{\rm i}\hat{\varSigma}^0_{11}\hat{g}_0-{\rm i}\hat{g}_0\hat{\varSigma}^0_{11}\nonumber \\
&+(\hat{\varDelta}_0+\hat{\varSigma}^0_{12})\hat{\bar{f}}_0
-\hat{f}_0(\hat{\varDelta}^\dag_0 -\hat{\varSigma}^0_{21})
=\hat{0},
\label{Eilenberger_ee}\\
\partial \hat{f}_0&+2\omega_n\hat{f}_0+{\rm i}\alpha \bm{g}_k \cdot (\hat{\bm{\sigma}}\hat{f}_0-\hat{f}_0\hat{\bm{\sigma}})
+{\rm i}\hat{\varSigma}^0_{11}\hat{f}_0+{\rm i}\hat{f}_0\hat{\varSigma}^0_{22}\nonumber \\
&+(\hat{\varDelta}_0+\hat{\varSigma}^0_{12})\hat{\bar{g}}_0
-\hat{g}_0(\hat{\varDelta}_0+\hat{\varSigma}^0_{12})
=\hat{0},
\label{Eilenberger_eh}\\
\partial \hat{\bar f}_0&-2\omega_n\hat{\bar f}_0+{\rm i}\alpha \bm{g}_k \cdot (\hat{\bm{\sigma}}\hat{\bar f}_0-\hat{\bar f}_0\hat{\bm{\sigma}})
-{\rm i}\hat{\varSigma}^0_{22}\hat{\bar f}_0-{\rm i}\hat{\bar f}_0\hat{\varSigma}^0_{11}\nonumber \\
&+(\hat{\varDelta}^\dag_0-\hat{\varSigma}^0_{21})\hat{g}_0
-\hat{\bar{g}}_0(\hat{\varDelta}^\dag_0 -\hat{\varSigma}^0_{21})
=\hat{0},
\label{Eilenberger_he}\\
\partial \hat{\bar g}_0&+{\rm i}\alpha \bm{g}_k \cdot (\hat{\bm{\sigma}}\hat{\bar g}_0-\hat{\bar g}_0\hat{\bm{\sigma}})
-{\rm i}\hat{\varSigma}^0_{22}\hat{\bar g}_0+{\rm i}\hat{\bar g}_0\hat{\varSigma}^0_{22}\nonumber \\
&+(\hat{\varDelta}^\dag_0-\hat{\varSigma}^0_{21})\hat{f}_0
-\hat{\bar{f}}_0(\hat{\varDelta}_0 +\hat{\varSigma}^0_{12})
=\hat{0},
\label{Eilenberger_hh}
\end{align}
\end{subequations}
with the normalization conditions
\begin{subequations}
\begin{align}
&\hat{g}^2_0+\hat{f}_0\hat{\bar{f}}_0=\hat{\sigma}_0,\label{normalization-a}\\
&\hat{g}_0\hat{f}_0+\hat{f}_0\hat{\bar{g}}_0=\hat{0},\label{normalization-b}\\
&\hat{\bar{f}}_0\hat{g}_0+\hat{\bar{g}}_0\hat{\bar{f}}_0=\hat{0},\label{normalization-c}\\
&\hat{\bar{f}}_0\hat{f}_0+\hat{\bar{g}}^2_0=\hat{\sigma}_0,\label{normalization-d}
\end{align}
\label{normalization-all}
\end{subequations}
where we define
$
\partial \equiv \bm{v}_{\rm F}\cdot \bm{\nabla},
\hat{g} \equiv\hat{g}_0,
\hat{f} \equiv\hat{f}_0{\rm i}\hat{\sigma}_y,
\hat{\bar{f}} \equiv-{\rm i}\hat{\sigma}_y\hat{\bar{f}}_0,
\hat{\bar{g}} \equiv-\hat{\sigma}_y\hat{\bar{g}}_0 \hat{\sigma}_y,
\hat{\varSigma}_{11} \equiv\hat{\varSigma}^0_{11},
\hat{\varSigma}_{12} \equiv\hat{\varSigma}^0_{12}{\rm i}\hat{\sigma}_y,
\hat{\varSigma}_{21} \equiv-{\rm i}\hat{\sigma}_y\hat{\varSigma}^0_{21},
\hat{\varSigma}_{22} \equiv-\hat{\sigma}_y\hat{\varSigma}^0_{22} \hat{\sigma}_y,
\hat{\varDelta} \equiv\hat{\varDelta}_0 {\rm i}\hat{\sigma}_y$,
and
$\hat{\varDelta}^\dag \equiv-{\rm i}\hat{\sigma}_y\hat{\varDelta}^\dag_0$.

\subsection{Impurity self-energy in the Born limit}
We next describe the nonmagnetic impurity scattering in a spatially uniform Rashba system.
We assume $s$-wave scattering in the Born limit.
The self-energy due to the nonmagnetic impurity scattering for split FSs is given by
\begin{equation}
\check{\varSigma}_{\rm I,II}({\rm i}\omega_n)=\frac{\varGamma_{\rm n}}{\pi}\langle \check{g}_{\rm I,II}({\rm i}\omega_n,\tilde{\bm{k}})\rangle_{\rm I,II},
\end{equation}
where $\varGamma_{\rm n}=\pi n_{\rm imp}N_{\rm F}v^2$ is the impurity scattering rate in the normal state,
$n_{\rm imp}$ is the density of impurities, $N_{\rm F}=(N_{\rm I}+N_{\rm II})/2$ is the DOS at the Fermi level in the normal state,
and $v$ is the $s$-wave scattering potential of an impurity.

We can separate the Green's functions with respect to each split band due to the ASOC using the band basis
where the normal-state Hamiltonian is diagonal (see the appendix of Ref. \cite{hayashi2006}),
at least in the case of spatially uniform systems \cite{frigeri2006} or spatially inhomogeneous systems within the clean limit ({\it e.g.}, clean vortex states) \cite{hayashi2006-vortex}.

Transformation of the Green's functions into the orbital basis (where the spin quantization axis is parallel to an applied field) then yields \cite{frigeri2006,hayashi2006,hayashi2006-vortex}
\begin{subequations}
\begin{align}
\hat{g}&=g_{\rm I}\hat{\sigma}_{\rm I}+g_{\rm II}\hat{\sigma}_{\rm II},\\
\hat{f}&=(f_{\rm I}\hat{\sigma}_{\rm I}+f_{\rm II}\hat{\sigma}_{\rm II}){\rm i}\hat{\sigma}_y,\\
\hat{\bar{f}}&=-{\rm i}\hat{\sigma}_y(\bar{f}_{\rm I}\hat{\sigma}_{\rm I}+\bar{f}_{\rm II}\hat{\sigma}_{\rm II}),\\
\hat{\bar{g}}&=-\hat{\sigma}_y(\bar{g}_{\rm I}\hat{\sigma}_{\rm I}+\bar{g}_{\rm II}\hat{\sigma}_{\rm II}){\rm i}\hat{\sigma}_y,
\end{align}
\end{subequations}
where $\hat{\sigma}_{\rm I,II}=(\hat{\sigma}_0\pm \bar{\bm{g}}_{\tilde{\bm{k}}}\cdot \hat{\bm{\sigma}})/2$
and $\bar{\bm{g}}_{\tilde{\bm{k}}}\equiv \bm{g}_{\tilde{\bm{k}}}/|\bm{g}_{\tilde{\bm{k}}}|$.
Provided that the Green's functions are invariant under the transformation $\tilde{\bm{k}} \rightarrow -\tilde{\bm{k}}$ \cite{note},
we obtain 
\begin{subequations}
\begin{align}
\langle\hat{g}\rangle_{\tilde{\bm{k}}}&=\frac{1}{2}\langle g_{\rm I}+g_{\rm II} \rangle_{\tilde{\bm{k}}}\hat{\sigma}_0,\\
\langle\hat{f}\rangle_{\tilde{\bm{k}}}&=\frac{1}{2}\langle f_{\rm I}+f_{\rm II} \rangle_{\tilde{\bm{k}}}\hat{\sigma}_0 {\rm i}\hat{\sigma}_y,\\
\langle\hat{\bar{f}}\rangle_{\tilde{\bm{k}}}&=-{\rm i}\hat{\sigma}_y \frac{1}{2}\langle \bar{f}_{\rm I}+\bar{f}_{\rm II} \rangle_{\tilde{\bm{k}}}\hat{\sigma}_0,\\
\langle\hat{\bar{g}}\rangle_{\tilde{\bm{k}}}&=-\hat{\sigma}_y\frac{1}{2}\langle \bar{g}_{\rm I}+\bar{g}_{\rm II} \rangle_{\tilde{\bm{k}}}\hat{\sigma}_0\hat{\sigma}_y,
\end{align}
\end{subequations}
because $\bar{\bm{g}}_{-\tilde{\bm{k}}}=-\bar{\bm{g}}_{\tilde{\bm{k}}}$.
Thus,
\begin{subequations}
\begin{align}
\langle \hat{g}_0 \rangle_{\tilde{\bm{k}}}       &= \cfrac{1}{2} \langle g_{\rm I} + g_{\rm II} \rangle_{\tilde{\bm{k}}} \hat{\sigma}_0,\\
\langle \hat{f}_0 \rangle_{\tilde{\bm{k}}}        &= \cfrac{1}{2} \langle f_{\rm I} + f_{\rm II} \rangle_{\tilde{\bm{k}}} \hat{\sigma}_0,\\
\langle \hat{\bar{f}}_0 \rangle_{\tilde{\bm{k}}} &= \cfrac{1}{2} \langle \bar{f}_{\rm I} + \bar{f}_{\rm II} \rangle_{\tilde{\bm{k}}} \hat{\sigma}_0,\\
\langle \hat{\bar{g}}_0 \rangle_{\tilde{\bm{k}}}  &= \cfrac{1}{2} \langle \bar{g}_{\rm I} + \bar{g}_{\rm II} \rangle_{\tilde{\bm{k}}} \hat{\sigma}_0.
\end{align}
\end{subequations}
Therefore,
\begin{subequations}
\begin{align}
&\hat{\varSigma}^0_{11} = -{\rm i}\varGamma_{\rm n} \langle \hat{g}_0 \rangle_{\tilde{\bm{k}}}=-{\rm i}\varGamma_{\rm n}\cfrac{1}{2}\langle g_{\rm I}+g_{\rm II} \rangle_{\tilde{\bm{k}}} \hat{\sigma}_0,
\label{impurity-self-energy11}\\
&\hat{\varSigma}^0_{12}  =   \varGamma_{\rm n} \langle \hat{f}_0 \rangle_{\tilde{\bm{k}}}=  \varGamma_{\rm n}\cfrac{1}{2}\langle f_{\rm I}+f_{\rm II} \rangle_{\tilde{\bm{k}}} \hat{\sigma}_0,
\label{impurity-self-energy12}\\
&\hat{\varSigma}^0_{21} =  -\varGamma_{\rm n} \langle \hat{\bar{f}}_0 \rangle_{\tilde{\bm{k}}}=  -\varGamma_{\rm n}\cfrac{1}{2}\langle \bar{f}_{\rm I}+\bar{f}_{\rm II} \rangle_{\tilde{\bm{k}}} \hat{\sigma}_0,
\label{impurity-self-energy21}\\
&\hat{\varSigma}^0_{22}  =  {\rm i}\varGamma_{\rm n} \langle \hat{\bar{g}}_0 \rangle_{\tilde{\bm{k}}}= {\rm i}\varGamma_{\rm n}\cfrac{1}{2}\langle \bar{g}_{\rm I}+\bar{g}_{\rm II} \rangle_{\tilde{\bm{k}}} \hat{\sigma}_0.
\label{impurity-self-energy22}
\end{align}
\end{subequations}
We note that, in the case of a spatially uniform system, the impurity self-energies $\hat{\varSigma}^0_{ij}$ are proportional to the unit matrix $\hat{\sigma}_0$.

Using Eqs.~(\ref{impurity-self-energy11})--(\ref{impurity-self-energy22}),
in the Eilenberger equations~(\ref{Eilenberger_ee}), (\ref{Eilenberger_eh}), (\ref{Eilenberger_he}), and (\ref{Eilenberger_hh}), respectively,
we obtain
\begin{subequations}
\begin{align}
{\rm i}\hat{\varSigma}^0_{11} \hat{g}_0-{\rm i}\hat{g}_0\hat{\varSigma}^0_{11}&=\hat{0},\\
2\omega_n \hat{f}_0+{\rm i}\hat{\varSigma}^0_{11}\hat{f}_0+{\rm i}\hat{f}_0\hat{\varSigma}^0_{22}
&=2(\omega_n+\sigma_{\rm g})\hat{\sigma}_0\hat{f}_0,\\
-2\omega_n \hat{\bar{f}}_0-{\rm i}\hat{\varSigma}^0_{22}\hat{\bar{f}}_0-{\rm i}\hat{\bar f}_0\hat{\varSigma}^0_{11}
&=-2(\omega_n+\sigma_{\rm g}) \hat{\sigma}_0\hat{\bar f}_0,\\
-{\rm i}\hat{\varSigma}^0_{22} \hat{\bar{g}}_0+{\rm i}\hat{\bar{g}}_0\hat{\varSigma}^0_{22}&=\hat{0},
\end{align}
\end{subequations}
where
\begin{align}
\sigma_{\rm g} &\equiv
\cfrac{\varGamma_{\rm n}}{2}(1+\delta) \cfrac{1}{2}\langle (g_{\rm I}-\bar{g}_{\rm I}) \rangle_{\tilde{\bm{k}}}+
\cfrac{\varGamma_{\rm n}}{2}(1-\delta) \cfrac{1}{2}\langle (g_{\rm II}-\bar{g}_{\rm II}) \rangle_{\tilde{\bm{k}}}\nonumber\\
&=
\frac{\varGamma_{\rm n}}{2}(1+\delta)\langle g_{\rm I} \rangle_{\tilde{\bm{k}}}+\frac{\varGamma_{\rm n}}{2}(1-\delta)\langle g_{\rm II}\rangle_{\tilde{\bm{k}}}.
\label{sigma_g}
\end{align}
In the spatially uniform system, we may use $g_{\rm I.II}+\bar{g}_{\rm I,II}=0$ to get Eq.~(\ref{sigma_g}).
The parameter $\delta=(N_{\rm I}-N_{\rm II})/2N_{\rm F}$ ($-1<\delta<1$) characterizes the difference in the DOSs between the split FSs I and II.
In the Eilenberger equations (\ref{Eilenberger_ee})--(\ref{Eilenberger_hh}), the rotation in spin space represented by the unitary matrix $\hat{U}_{\tilde{\bm{k}}}$ yields
\begin{align}
\hat{U}^\dag_{\tilde{\bm{k}}} {\rm i}\alpha \bm{g}_{\tilde{\bm{k}}} \cdot(\hat{\bm{\sigma}}\hat{A}_0-\hat{A}_0\hat{\bm{\sigma}})&\hat{U}_{\tilde{\bm{k}}}
=2{\rm i}\alpha|\bm{g}_{\tilde{\bm{k}}}|
\left(
\begin{array}{cc}
0 & -A_{\rm b}\\
 A_{\rm c} & 0
\end{array}
\right),\\
\hat{U}^\dag_k(\hat{\varDelta_0}+\hat{\varSigma}^0_{12})\hat{U}_{\tilde{\bm{k}}}
=&
\left(
\begin{array}{cc}
\varDelta_{\rm II}+\sigma_{\rm f} & 0 \\
0 & \varDelta_{\rm I}+\sigma_{\rm f}
\end{array}
\right),\\
\hat{U}^\dag_{\tilde{\bm{k}}}(\hat{\varDelta_0}^\dag-\hat{\varSigma}^0_{21})\hat{U}_{\tilde{\bm{k}}}
=&
\left(
\begin{array}{cc}
\varDelta^\ast_{\rm II}+\bar{\sigma}_{\rm f} & 0 \\
0 & \varDelta^\ast_{\rm I}+\bar{\sigma}_{\rm f}
\end{array}
\right),
\end{align}
where $\hat{A}_0$ refers to $\hat{g}_0$, $\hat{f}_0$, $\hat{\bar{f}}_0$, or $\hat{\bar{g}}_0$ and
\begin{align}
&\hat{U}^\dag_{\tilde{\bm{k}}} \hat{A}_0 \hat{U}_{\tilde{\bm{k}}}
\equiv
\left(
\begin{array}{cc}
A_{\rm a} & A_{\rm b} \\
A_{\rm c} & A_{\rm d}
\end{array}
\right),\\
&\varDelta_{\rm I,II} \equiv \psi_{\rm s} \pm d_{\rm t}|\bm{g}_{\tilde{\bm{k}}}|,\\
&\sigma_{\rm f} \equiv \varGamma_{\rm n}(1+\delta) \cfrac{1}{2} \langle f_{\rm I} \rangle_{\tilde{\bm{k}}}+\varGamma_{\rm n}(1-\delta) \cfrac{1}{2} \langle f_{\rm II} \rangle,
\label{sigma_f}
\\
&\bar{\sigma}_{\rm f} \equiv \varGamma_{\rm n}(1+\delta) \cfrac{1}{2} \langle \bar{f}_{\rm I} \rangle_{\tilde{\bm{k}}}+\varGamma_{\rm n}(1-\delta) \cfrac{1}{2} \langle \bar{f}_{\rm II} \rangle.
\end{align}
Hence,
the impurity effect in the spatially uniform state appears only in the replacement of the Matsubara frequency and the order parameter:
\begin{align}
\omega_n &\rightarrow \tilde{\omega}_n\equiv\omega_n+\sigma_{\rm g},\\
\varDelta_{\rm I,II} &\rightarrow \tilde{\varDelta}_{\rm I,II}\equiv \varDelta_{\rm I,II}+\sigma_{\rm f},\\
\varDelta^\ast_{\rm I,II} &\rightarrow \tilde{\varDelta}^\ast_{\rm I,II}\equiv \varDelta_{\rm I,II}+\bar{\sigma}_{\rm f}.
\end{align}

\subsection{Gap equation}
In the same manner as Ref.~\cite{hayashi2006},
we obtain the following Eilenberger equations in the band basis in the presence of impurities for the suffix a:
\begin{subequations}
\begin{align}
&\tilde{\varDelta}_{\rm II}\bar{f}_{\rm a}-\tilde{\varDelta}^\ast_{\rm II}f_{\rm a}=0,\\
&2\tilde{\omega}_n f_{\rm a}+\tilde{\varDelta}_{\rm II}\bar{g}_{\rm a}-\tilde{\varDelta}_{\rm II}g_{\rm a}=0,\\
&2\tilde{\omega}_n\bar{f}_{\rm a}-\tilde{\varDelta}^\ast_{\rm II}g_{\rm a}+\tilde{\varDelta}^\ast_{\rm II}\bar{g}_{\rm a}=0,\\
&\tilde{\varDelta}^\ast_{\rm II}f_{\rm a} -\tilde{\varDelta}_{\rm II}\bar{f}_{\rm a}=0.
\end{align}
\label{eilenberger-a}
\end{subequations}
For the suffix b, we obtain
\begin{subequations}
\begin{align}
&-2{\rm i}\alpha|\bm{g}_{\tilde{\bm{k}}}|  g_{\rm b}+\tilde{\varDelta}_{\rm II}\bar{f}_{\rm b}-\tilde{\varDelta}^\ast_{\rm I}f_{\rm b}=0,\\
&2\tilde{\omega}_n f_{\rm b}-2{\rm i}\alpha|\bm{g}_{\tilde{\bm{k}}}| f_{\rm b} +\tilde{\varDelta}_{\rm II} \bar{g}_{\rm b}-\tilde{\varDelta}_{\rm I}g_{\rm b}=0,\\
&2\tilde{\omega}_n\bar{f}_{\rm b}+2{\rm i}\alpha|\bm{g}_{\tilde{\bm{k}}}| \bar{f}_{\rm b} -\tilde{\varDelta}^\ast_{\rm II} g_{\rm b}+\tilde{\varDelta}^\ast_{\rm I}\bar{g}_{\rm b}=0,\\
&-2{\rm i}\alpha|\bm{g}_{\tilde{\bm{k}}}| \bar{g}_{\rm b}+\tilde{\varDelta}^\ast_{\rm II}f_{\rm b}-\tilde{\varDelta}_{\rm I}\bar{f}_{\rm b}=0.
\end{align}
\end{subequations}
For the suffix c, we obtain
\begin{subequations}
\begin{align}
&2{\rm i}\alpha|\bm{g}_{\tilde{\bm{k}}}| g_{\rm c} + \tilde{\varDelta}_{\rm I}\bar{f}_{\rm c}-\tilde{\varDelta}^\ast_{\rm II}f_{\rm c}=0,\\
&2\tilde{\omega}_n f_{\rm c} + 2{\rm i}\alpha|\bm{g}_{\tilde{\bm{k}}}| f_{\rm c} +\tilde{\varDelta}_{\rm I} \bar{g}_{\rm c}-\tilde{\varDelta}_{\rm II}g_{\rm c}=0,\\
&2\tilde{\omega}_n\bar{f}_{\rm c}-2{\rm i}\alpha|\bm{g}_{\tilde{\bm{k}}}| \bar{f}_{\rm c} -\tilde{\varDelta}^\ast_{\rm I} g_{\rm c}+\tilde{\varDelta}^\ast_{\rm II}\bar{g}_{\rm c}=0,\\
&2{\rm i}\alpha|\bm{g}_{\tilde{\bm{k}}}| \bar{g}_{\rm c}+\tilde{\varDelta}^\ast_{\rm I}f_{\rm c}-\tilde{\varDelta}_{\rm II}\bar{f}_{\rm c}=0.
\end{align}
\end{subequations}
Finally, for the suffix d, we obtain
\begin{subequations}
\begin{align}
&\tilde{\varDelta_{\rm I}}\bar{f}_{\rm d}-\tilde{\varDelta}^\ast_{\rm I}f_{\rm d}=0,\\
&2\tilde{\omega}_n f_{\rm d}+\tilde{\varDelta}_{\rm I}\bar{g}_{\rm d}-\tilde{\varDelta}_{\rm I}g_{\rm d}=0,\\
&2\tilde{\omega}_n\bar{f}_{\rm d}-\tilde{\varDelta}^\ast_{\rm I}g_{\rm d}+\tilde{\varDelta}^\ast_{\rm I}\bar{g}_{\rm d}=0,\\
&\tilde{\varDelta}^\ast_{\rm I}f_{\rm d} -\tilde{\varDelta}_{\rm I}\bar{f}_{\rm d}=0.
\end{align}
\label{eilenberger-d}
\end{subequations}
As discussed in Ref.~\cite{hayashi2006},
we note that $g_{\rm b,c}=f_{\rm b,c}=\bar{f}_{\rm b,c}=\bar{g}_{\rm b,c}=0$ for $\alpha|\bm{g}_{\tilde{\bm{k}}}|\neq 0$.
However, as opposed to the clean-limit case, this result is valid only for spatially uniform systems.
By transforming the normalization condition (\ref{normalization-b}) or (\ref{normalization-c}) into that in the band basis,
we obtain $g_{\rm a,d}=-\bar{g}_{\rm a,d}$ for spatially uniform systems.
From this relation, the normalization condition (\ref{normalization-a}), and the Eilenberger equations (\ref{eilenberger-a}) and (\ref{eilenberger-d}),
we obtain the following Green's functions for spatially uniform systems:
\begin{subequations}
\begin{align}
g_{\rm d,a}&\equiv g_{\rm I,II}({\rm i}\omega_n,\tilde{\bm{k}})=\cfrac{\omega_n+\sigma_{\rm g}({\rm i}\omega_n)}{\varLambda_{\rm I,II}({\rm i}\omega_n,\tilde{\bm{k}})},\\
f_{\rm d,a}&\equiv f_{\rm I,II}({\rm i}\omega_n,\tilde{\bm{k}})=\cfrac{\varDelta_{\rm I,II}(\tilde{\bm{k}})+\sigma_{\rm f}({\rm i}\omega_n)}{\varLambda_{\rm I,II}({\rm i}\omega_n,\tilde{\bm{k}})},\\
\bar{f}_{\rm d,a}&\equiv \bar{f}_{\rm I,II}({\rm i}\omega_n,\tilde{\bm{k}})=\cfrac{\varDelta^\ast_{\rm I,II}(\tilde{\bm{k}})+\bar{\sigma}_{\rm f}({\rm i}\omega_n)}{\varLambda_{\rm I,II}({\rm i}\omega_n,\tilde{\bm{k}})},\\
\bar{g}_{\rm d,a}&\equiv \bar{g}_{\rm I,II}({\rm i}\omega_n,\tilde{\bm{k}})=\cfrac{-\omega_n-\sigma_{\rm g}({\rm i}\omega_n)}{\varLambda_{\rm I,II}({\rm i}\omega_n,\tilde{\bm{k}})},
\end{align}
\end{subequations}
where $\varLambda_{\rm I,II}({\rm i}\omega_n,\tilde{\bm{k}})=\sqrt{\tilde{\omega}^2_n +|\tilde{\varDelta}_{\rm I,II}|^2}$.
The corresponding gap equations are
\begin{eqnarray}
\psi_{\rm s}
&=&
\pi T\sum_{|\omega_n|<\omega_{\rm c}}
\bigl[
\lambda_{\rm s}\bigl\{\langle f_+\rangle_0+\delta\langle f_-\rangle_0  \bigr\}
\nonumber\\
&&+\lambda_{\rm m}\bigl\{ \langle |\bm{g}_{\tilde{\bm{k}}}|f_- \rangle_0+\delta\langle |\bm{g}_{\tilde{\bm{k}}}|f_+ \rangle_0 \bigr\}
\bigr],
\label{gap-eq1_isotropic}
\\
d_{\rm t}
&=&
\pi T\sum_{|\omega_n|<\omega_{\rm c}}
\bigl[
\lambda_{\rm t}\bigl\{\langle |\bm{g}_{\tilde{\bm{k}}}|f_- \rangle_0+\delta \langle |\bm{g}_{\tilde{\bm{k}}}|f_+ \rangle_0  \bigr\}
\nonumber\\
&&+\lambda_{\rm m}\bigl\{ \langle f_+ \rangle_0+\delta\langle f_- \rangle_0 \bigr\}
\bigr],
\label{gap-eq2_isotropic}
\end{eqnarray}
where $f_\pm\equiv(f_{\rm I}+f_{\rm II})/2$ and $\omega_{\rm c}$ is the cutoff frequency.
In the clean limit and in the limit of $T\rightarrow T_{\rm c}$,
the coupling constants are determined as follows \cite{hayashi2006}:
\begin{align}
\lambda_{\rm s}
&=\cfrac{\lambda \nu-\lambda_{\rm m}( 1+\delta \nu \langle |\bm{g}_{\tilde{\bm{k}}}|\rangle_0)}{ \nu + \delta\langle |\bm{g}_{\tilde{\bm{k}}}|\rangle_0},\\
\lambda_{\rm t}
&=\cfrac{\lambda-\lambda_{\rm m}( \nu+\delta \langle |\bm{g}_{\tilde{\bm{k}}}|\rangle_0) }{1+\delta\nu\langle |\bm{g}_{\tilde{\bm{k}}}| \rangle_0 }.
\end{align}
Here, $\nu$, $\lambda$, and $n_{\rm c}(T)$ are defined in the main text.

\section{$H^{||}_{\rm c2}(T)$ variation with parity mixing}
\label{appendixC}
\begin{figure}[tb]
  \begin{center}
    \begin{tabular}{p{85mm}}
      \resizebox{85mm}{!}{\includegraphics{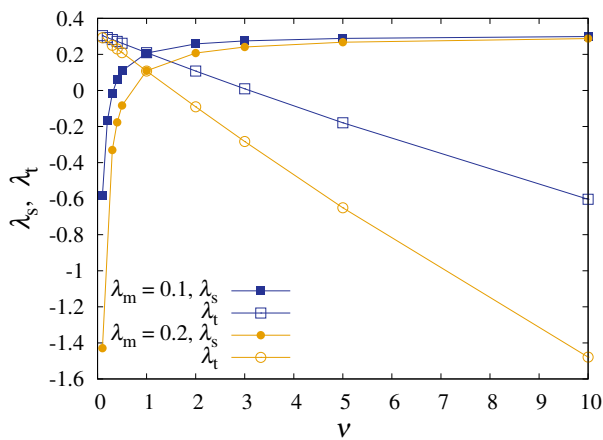}}
    \end{tabular}
\caption{
\label{fig1_appx}
(Color online)
Mixing ratio dependence of the coupling constants for the singlet and triplet channels evaluated for Si(111)-($\sqrt{7}\times\sqrt{3}$)-In.
The DOS difference between the split FSs is set to $\delta=0$.}
  \end{center}
\end{figure}
\begin{figure*}[tb]
    \begin{tabular}{p{90mm}p{90mm}p{90mm}p{90mm}}
      \resizebox{90mm}{!}{\includegraphics{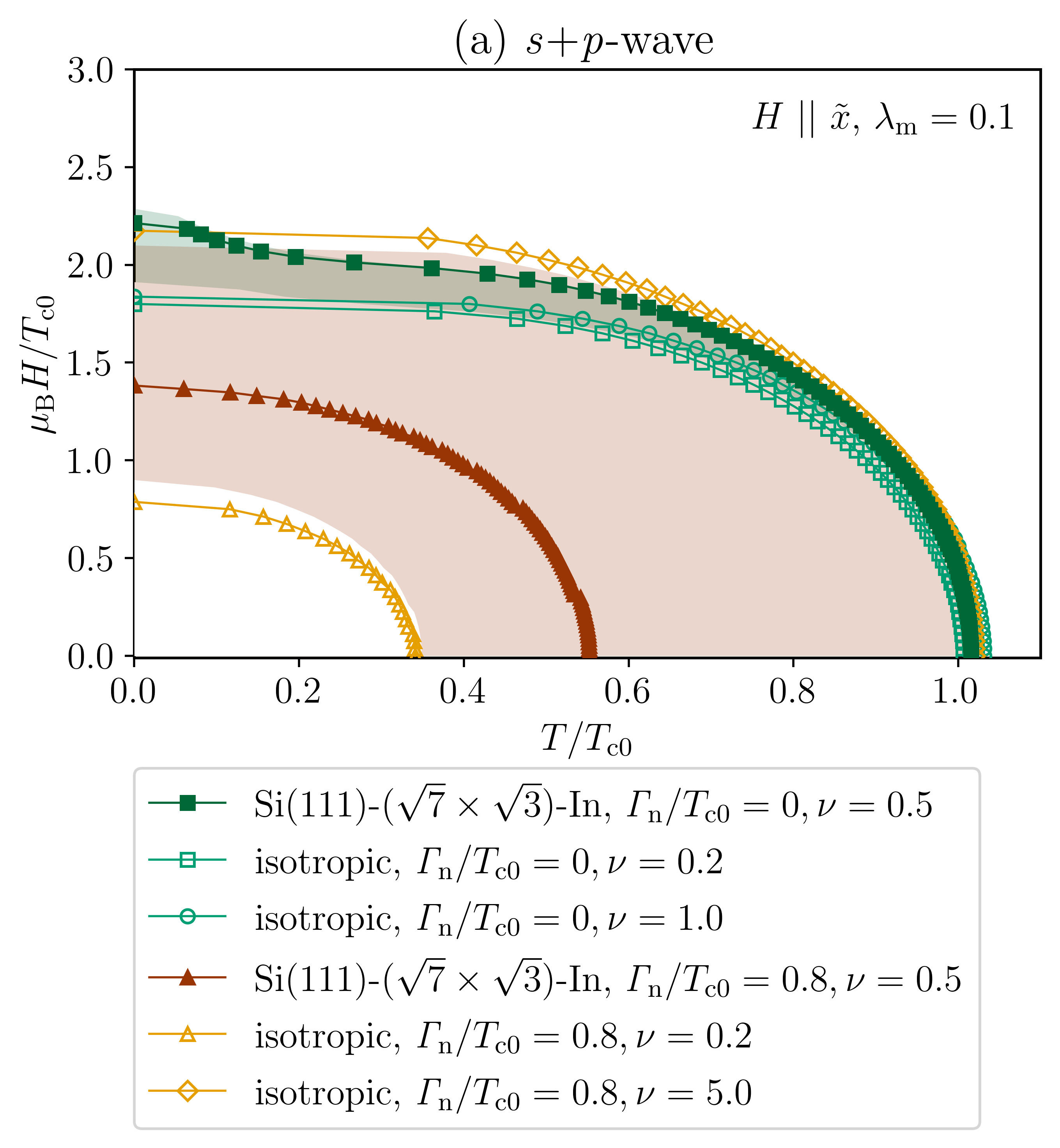}}&
      \resizebox{90mm}{!}{\includegraphics{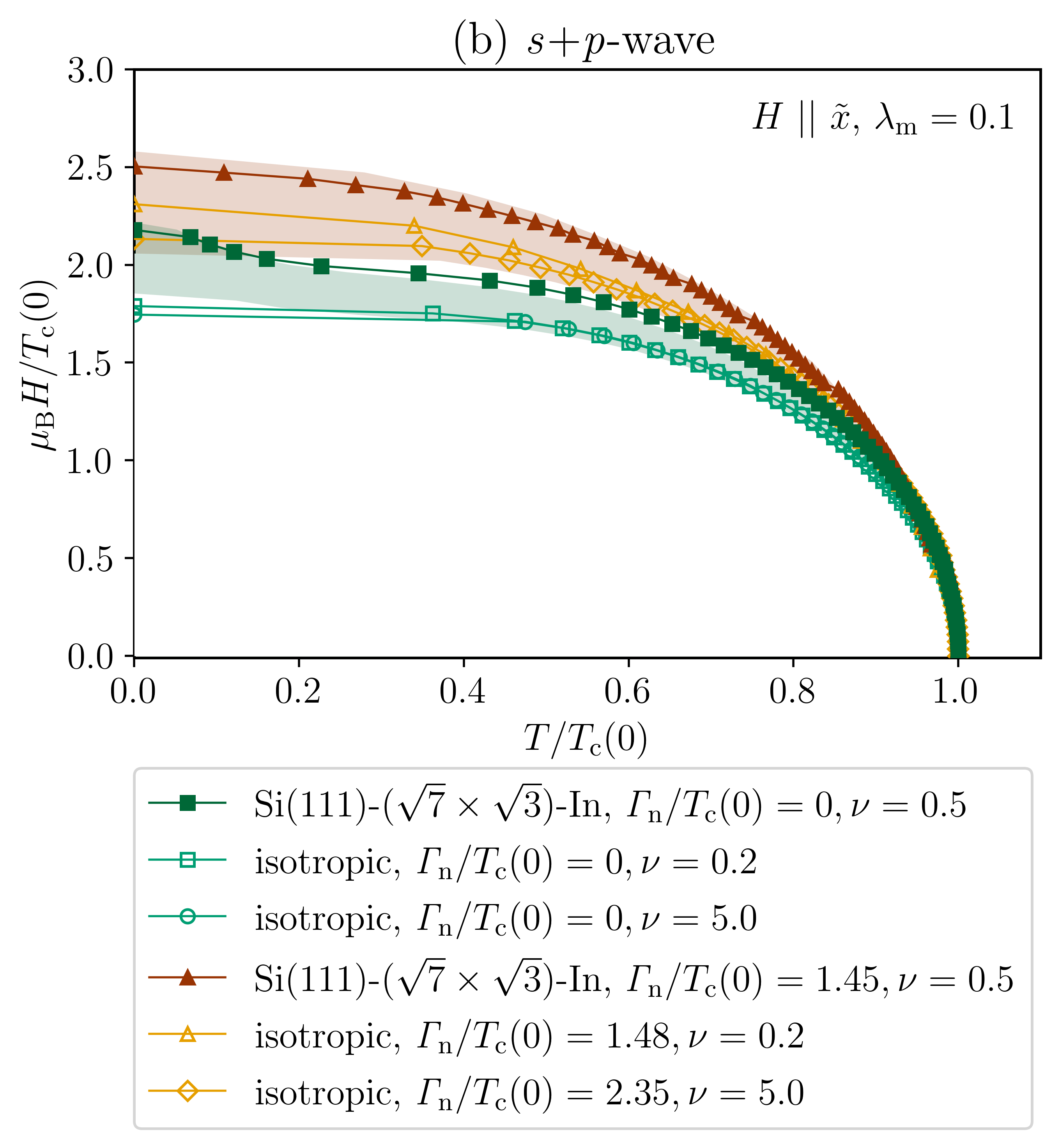}}\\
      \resizebox{90mm}{!}{\includegraphics{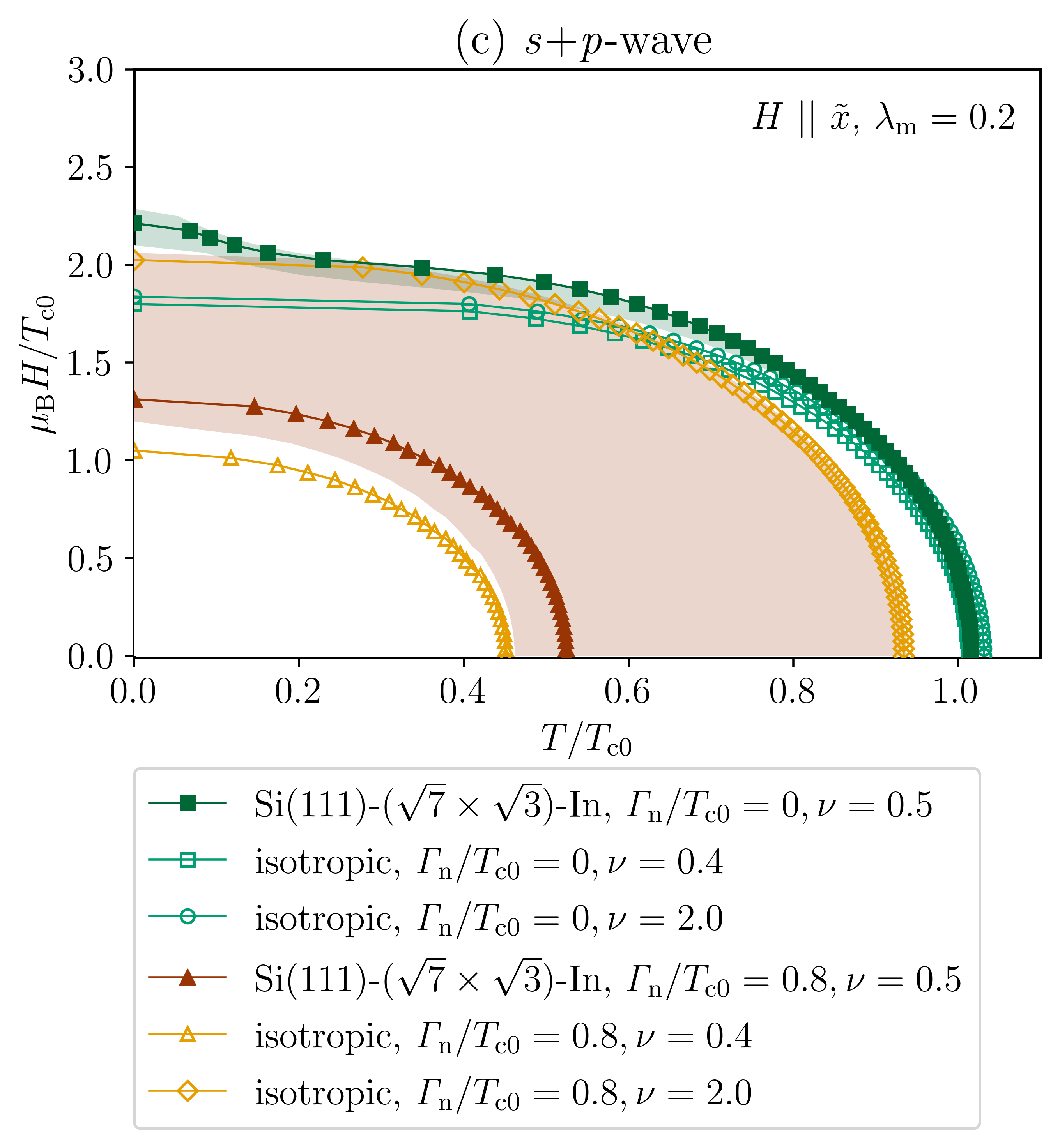}}&
      \resizebox{90mm}{!}{\includegraphics{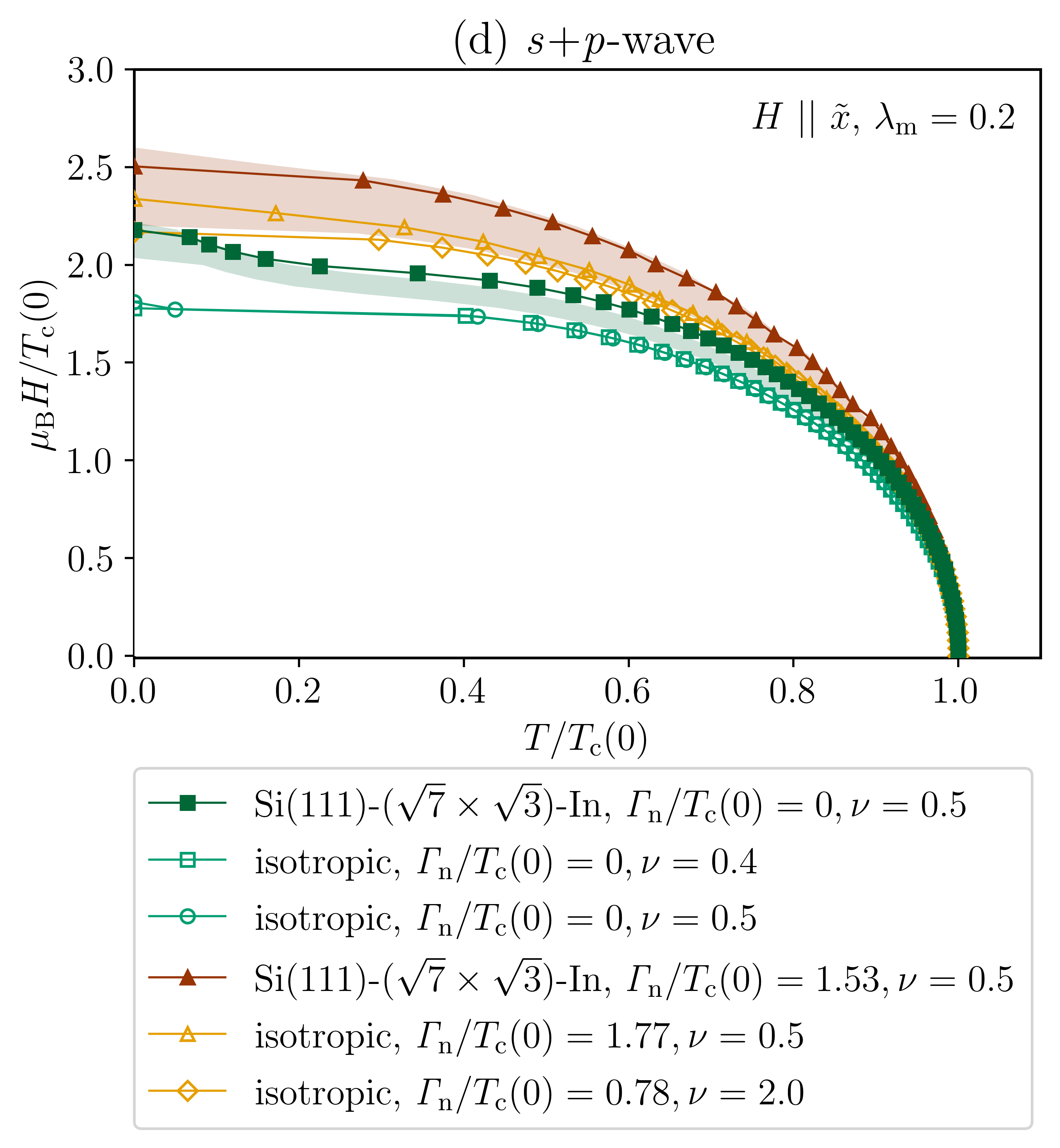}}
    \end{tabular}
\caption{
\label{fig2_appx}
(Color online)
Temperature dependence of an in-plane critical magnetic field oriented parallel to the $x$ axis for the $s$+$p$-wave pairing upon changing the normal state scattering rate $\varGamma_{\rm n}$ and the parity mixing ratio $\nu$ for (a), (b) $\lambda_{\rm m}=0.1$ and (c), (d) $\lambda_{\rm m}=0.2$.
The filled and open symbols denote the data for the Fermi surface of a Si(111)-$(\sqrt{7}\times\sqrt{3})$-In and an isotropic system, respectively.
The difference of the density of states between the split two FSs is set to $\delta=0$.
Temperature and magnetic field are normalized by (a), (c) $T_{\rm c0}$ and (b), (d) $T_{\rm c}(0)$, respectively.
}
\end{figure*}
\begin{figure*}[tb]
    \begin{tabular}{p{90mm}p{90mm}p{90mm}p{90mm}}
      \resizebox{90mm}{!}{\includegraphics{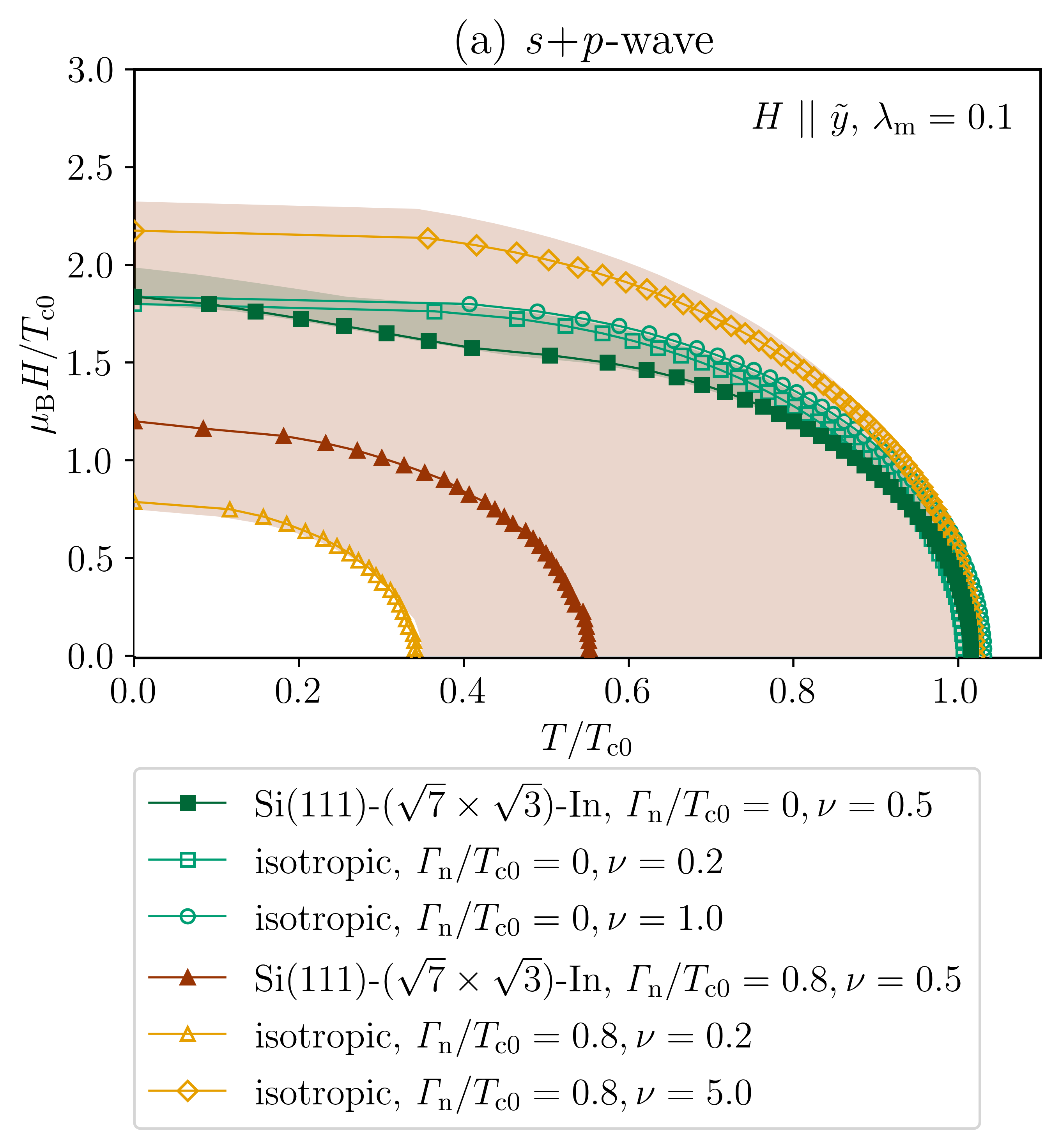}}&
      \resizebox{90mm}{!}{\includegraphics{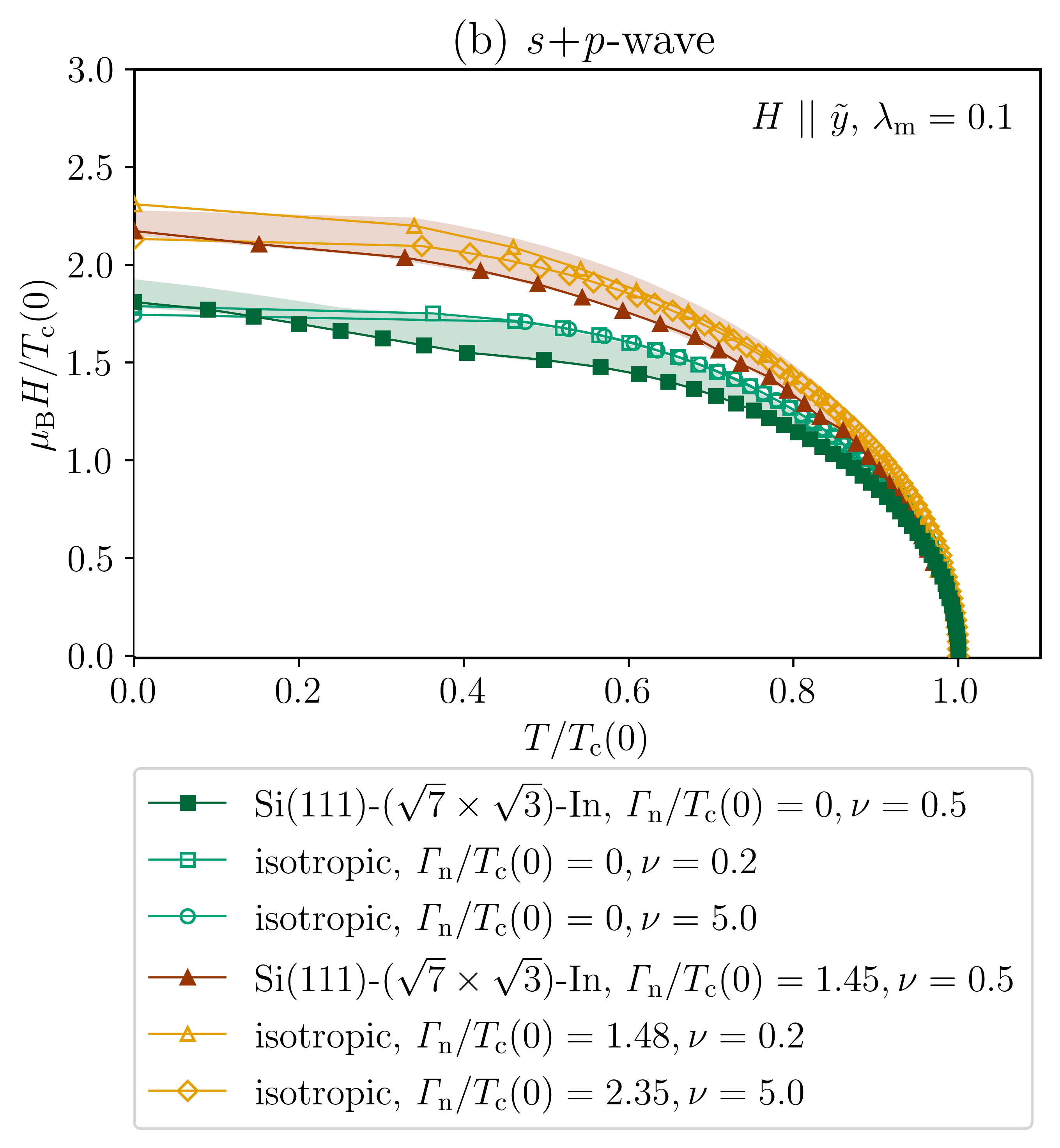}}\\
      \resizebox{90mm}{!}{\includegraphics{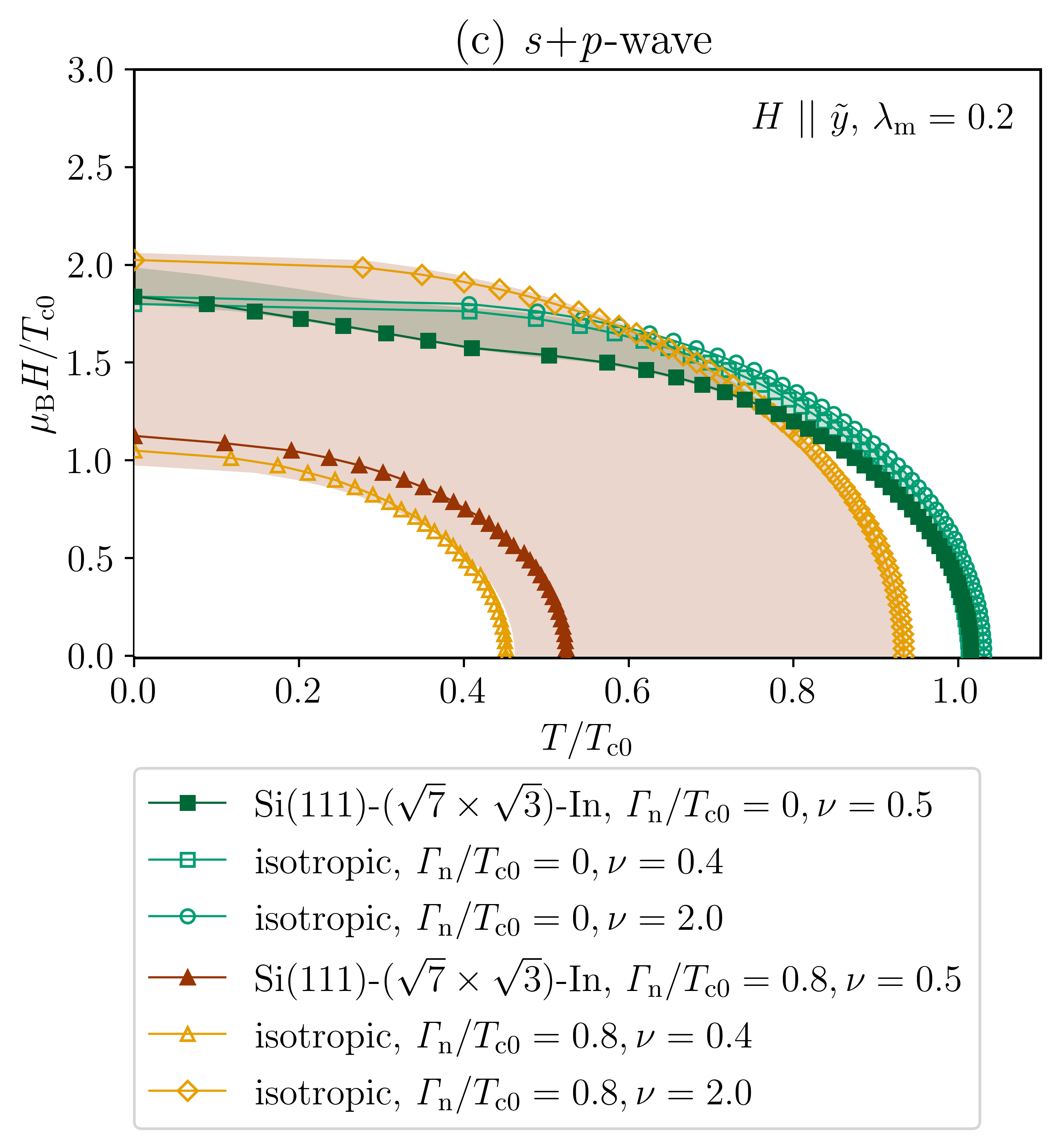}}&
      \resizebox{90mm}{!}{\includegraphics{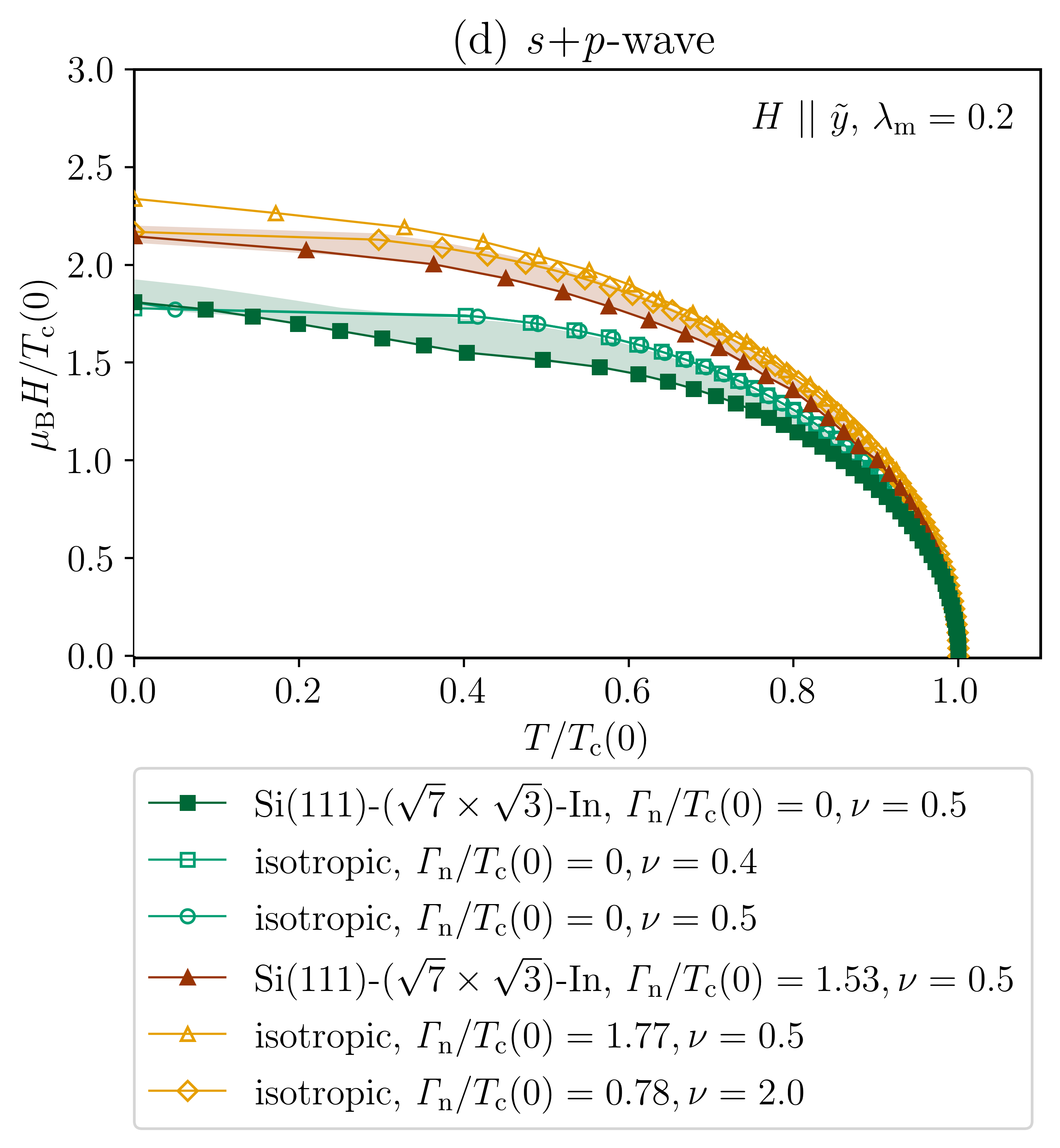}}
    \end{tabular}
\caption{
\label{fig3_appx}
(Color online)
Temperature dependence of an in-plane critical magnetic field oriented parallel to the $y$ axis for the $s$+$p$-wave pairing upon changing the normal state scattering rate $\varGamma_{\rm n}$ and the parity mixing ratio $\nu$ for (a), (b) $\lambda_{\rm m}=0.1$ and (c), (d) $\lambda_{\rm m}=0.2$.
The filled and open symbols denote the data for the Fermi surface of a Si(111)-$(\sqrt{7}\times\sqrt{3})$-In and an isotropic system, respectively.
The difference of the density of states between the split two FSs is set to $\delta=0$.
Temperature and magnetic field are normalized by (a), (c) $T_{\rm c0}$ and (b), (d) $T_{\rm c}(0)$, respectively.
}
\end{figure*}
To study the variation of $H^{||}_{\rm c2}(T)$ when $\nu$ changes over a wide range,
it turned out that the coupling constant for the mixing channel is limited to $\lambda_{\rm m}\ll1$.
This is because the self-consistent solutions of the order parameter can be obtained only when $-0.2 \lesssim \lambda_{\rm s}, \lambda_{\rm t}\lesssim 0.3$ (see Fig. \ref{fig1_appx}),
suggesting that the variation range of $\nu$ narrows as $\lambda_{\rm m}$ increases.
Therefore, we will limit ourselves to $\lambda_{\rm m}=0.1$ and $0.2$. 
Fig. \ref{fig1_appx} shows the dependence of the coupling constants $\lambda_{\rm s}$ and $\lambda_{\rm t}$ on $\nu$
evaluated by Eqs. (\ref{coupling_2band_singlet}) and (\ref{coupling_2band_triplet}) for Si(111)-($\sqrt{7}\times\sqrt{3}$)-In.
The coupling constants are controlled by the input parameters $\lambda_{\rm m}$ and $\nu$.
At the equal mixing ratio ($\nu=1$), $\lambda_{\rm s}\approx \lambda_{\rm t}$ for $\lambda_{\rm m}=0.1$ and $0.2$, respectively.
For $\lambda_{\rm m}=0.1$, the self-consistent solution of the order parameter was not obtained for $\nu=0.1$ and $10$,
while for $\lambda_{\rm m}=0.2$, it was not obtained for $\nu \lesssim 0.3$ or $\nu \gtrsim 3$.
We consider that the choice of $\lambda_{\rm m}=0.1$ and $0.2$ is reasonable, since these values correspond to the weak coupling regime.

In Figs. \ref{fig2_appx} and \ref{fig3_appx}, we show the $H^{||}_{\rm c2}(T)$ oriented parallel to the $x$ and $y$ axes, respectively, in the case of the $s$+$p$-wave paring
to check the stability of the results in Figs. \ref{fig4} and \ref{fig5}, respectively, upon changing $\nu$ and $\lambda_{\rm m}$.
The physical quantities are rescaled by $T_{\rm c}(0)$ in Figs. \ref{fig2_appx}(b), \ref{fig2_appx}(d), \ref{fig3_appx}(b), and \ref{fig3_appx}(d).
The green and red shaded areas in Figs. \ref{fig2_appx} and \ref{fig3_appx} depict the variation range of the transition line when $\nu$ varies in the case of $\varGamma_{\rm n}/T_{\rm c0}=0$ and $0.8$, respectively for Si(111)-($\sqrt{7}\times\sqrt{3}$)-In.
For an isotropic FS, the variation range of $H^{||}_{\rm c2}$ is shown by the area between open symbols.
For $\lambda_{\rm m}=0.1~(0.2)$, we varied $\nu$ in the range $0.2\le\nu\le5~(0.4\le\nu\le2)$ in any conditions to show the variation of the transition line such that the variation of $H^{||}_{\rm c2}(0)$ is the largest.
 
Figure \ref{fig2_appx} shows the same trend of $H^{||}_{\rm c2}(T\approx 0)$ for Si(111)-($\sqrt{7}\times\sqrt{3}$)-In being larger than that for isotropic systems as in Fig. \ref{fig4}.
Fig. \ref{fig3_appx} also shows the same trend of the suppression of the $H^{||}_{\rm c2}(T\approx 0)$ enhancement as in Fig. \ref{fig5},
although it shows some variation depending on $\nu$.


\bibliography{reference-2D-Rashba-imp-uniform}

\begin{thebibliography}{65}
\bibitem [{mem()}]{memo}%
  \BibitemOpen
  \href@noop {} {}\bibinfo {note} {The slight discrepancy between $T_{\rm
  c}(0)$ and $T_{\rm c0}$ stems from the difference in their definition.
  $T_{\rm c0}$ is the $T_{\rm c}$ for an $s$-wave pair in the clean limit in
  the absence of the ASOC and magnetic field. $T_{\rm c0}$ is the temperature
  satisfying $1/\lambda=\sum_{n=0}^{n_{\rm c}(T_{\rm c0})}1/(n+1/2)$, from
  which we estimate $\lambda\approx 0.298$. On one hand, $T_{\rm c}(0)$ denotes
  the $T_{\rm c }$ at $H=0$. In the simplest case ($s$-wave pair, isotropic
  Fermi surface, and $\delta=0$), $T_{\rm c}(0)$ is determined by solving
  $\psi_{\rm s}=\lambda \pi T \sum_{|\omega_n|<\omega_{\rm c}}\psi_{\rm
  s}/\sqrt{\omega^2_n+\psi^2_{\rm s}}$. We estimate $\lambda\approx 0.323$ via
  $1/\lambda=\sum_{n=0}^{n_{\rm c}(T)}1/(n+1/2)+\ln(T/T_{\rm c0})$ for the
  self-consistent solution of $T_{\rm c}(0)/T_{\rm c0}=1.031$. For both
  estimates, we use the cutoff frequency $\omega_{\rm c}=7\pi T_{\rm c0}$. The
  fact that $T_{\rm c}(0)/T_{\rm c0}>1$ reflects the value of the coupling
  constant used to determine $T_{\rm c}(0)$.}\BibitemShut {Stop}%
\end{thebibliography}

\end{document}